\input eplain

\def\toprule{\vskip1.5pt\hrule height0.8pt\vskip1.5pt}
\def\midrule{\vskip1.5pt\hrule\vskip1.5pt}
\def\bottomrule{\vskip1.5pt\hrule height0.8pt\vskip1.5pt}

\def\nomenclature[#1]#2#3{\line{{#2\hfill} \hbox to0.85\hsize {#3\hfill}}}

\newcount\fignumber
\def\figdef#1{\global\advance\fignumber by 1 \definexref{#1}{\number\fignumber}{figure}\ref{#1}}
\def\figdefn#1{\global\advance\fignumber by 1 \definexref{#1}{\number\fignumber}{figure}}
\let\figref=\ref
\let\figrefn=\refn
\let\figrefs=\refs

\newcount\tabnumber
\def\tabdef#1{\global\advance\tabnumber by 1 \definexref{#1}{\number\tabnumber}{table}\ref{#1}}
\def\tabdefn#1{\global\advance\tabnumber by 1 \definexref{#1}{\number\tabnumber}{table}}
\let\tabref=\ref

\input epsf

\def\figscale#1#2{\epsfxsize=#2\epsfbox{#1.eps}}

\newcount\scount \scount=0
\newcount\sscount \sscount=0

\makeatletter
\def\section#1\par{
  \vskip\z@ plus.3\vsize\penalty-250
  \vskip\z@ plus-.3\vsize\bigskip\vskip\parskip
  \global\advance\scount by1
  \sscount=0
  \writenumberedtocentry{section}#1{\the\scount}
  \definexref#1{\the\scount}{section}
  \message{#1}
  \noindent\the\scount.\quad{\bf #1}\nobreak\smallskip\noindent}
\makeatother

\def\subsection#1{
  \global\advance\sscount by1
  \smallskip
  \noindent{~~\the\scount.\the\sscount~~{\bf{#1.~}}}}

\def\Gr{{Gr}}
\def\Ls{{L\!^*}}

\def\NuolR{{\overline{N\!u_R}}}
\def\Nuolq{\overline{N\!u'}}

\def\Nuolb{\overline{N\!u^\bullet}}
\def\Nuolp{\overline{N\!u^\|}}
\def\holb{\overline{h^\bullet}}
\def\holp{\overline{h^\|}}
\def\holz{\overline{h_0}}

\def\Nuols{\overline{N\!u^*}}
\def\Nuol{{\overline{N\!u}}}
\def\Shol{{\overline{Sh}}}

\def\Nuzq{{N\!u'_0}}
\def\Nuzs{{N\!u^*_0}}
\def\Nuzb{{N\!u^\bullet_0}}
\def\Nuz{{N\!u_0}}
\def\Nu{{N\!u}}
\def\Pra{{Pr}}
\def\Sc{{Sc}}
\def\Ra{{Ra}}

\def\diff{{\rm d}}

\def\hRol{{\overline{h_R}}}
\def\hol{{\overline{h}}}
\def\hqol{{\overline{h'}}}

\def\hsol{{\overline{h^*}}}

\def\huol{{\overline{h_\uparrow}}}
\def\hdol{{\overline{h_\downarrow}}}

\def\etal{{et~al.~}}

\def\Uol{\overline{U}}

\centerline{\bf{Natural Convection Heat Transfer from an Inclined Cylinder}}
\medskip
\centerline{Aubrey G. Jaffer and Martin S. Jaffer}
\centerline{e-mail: agj@alum.mit.edu}

\beginsection{Abstract}

{\narrower

 Based on Jaffer's (2023) heat engine analysis of natural convection,
 this investigation mathematically derives a novel, comprehensive
 formula predicting the natural convective heat transfer from an
 inclined cylinder given its length, diameter, angle, and Rayleigh
 number, and the fluid's Prandtl number and thermal conductivity.

 The present formula was tested with 116 inclined cylinder
 measurements having length-to-diameter ratios between 1.48 and 12500
 in ten data-sets from four peer-reviewed studies, yielding (data-set)
 root-mean-squared relative error values between 1.5\% and 5\%.

\par}

\medskip
{\noindent {\bf Keywords}: natural-convection; cylinder; inclined}

\medskip
This research did not receive any specific grant from funding agencies
in the public, commercial, or not-for-profit sectors.

\beginsection{Table of Contents}

\readtocfile

\bigskip
\vbox{\settabs 2\columns
\+\hfil\figscale{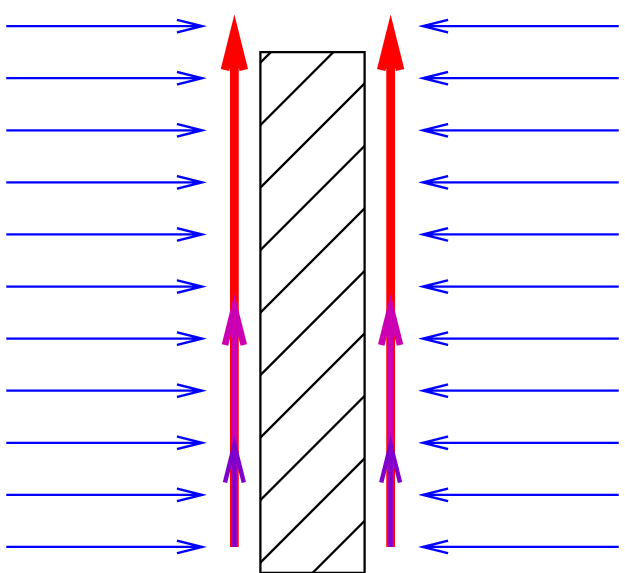}{160pt} & \hfil\figscale{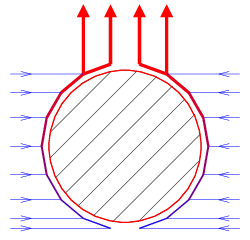}{160pt}&\cr
\+\hfil{\bf~a}&\hfil{\bf~b}&\cr
}
\vbox{\settabs 1\columns
\+\hfil{\bf\figdef{fig:vertical-flow}\quad{Induced flow around ({\bf a}) vertical cylinder and ({\bf b}) level cylinder.}}&\cr
}
\medskip

\section{Introduction}

  Natural convection is the flow caused by nonuniform density in a
  fluid under the influence of gravity.
  Natural convection is a fundamental process with application from
  engineering to geophysics.

  Changes in fluid density can be caused by changes in temperature or
  solute concentration.  Under the influence of gravity, density
  changes cause fluid flow, which also transports heat or solute.
  Rates of transfer grow until reaching a plateau.
  This investigation seeks to predict the overall steady-state heat
  transfer rate from an external, cylindrical, isothermal surface
  inclined at any angle in a Newtonian fluid.

  An ``external'' surface is one that fluid can flow around freely,
  especially horizontally.  If enclosed, the enclosure must have
  dimensions much larger than the heated or cooled surface.

 Unbounded vertical extent allows the imagining of a hypothetical
 apparatus to extract as much of the rising fluid's mechanical energy
 as possible.  This then is a (non-reversible) heat engine driven by
 the temperature difference between the heated object and the unheated
 fluid.
 The analysis in Jaffer~\cite{thermo3010010} finds that the maximum
 efficiency, the fraction of the heat energy which can converted into
 mechanical work, is $|T-T_\infty|/[2\,T_\infty]$, which is 1/2 of the
 Carnot efficiency limit for reversible heat engines.

 From this thermodynamic constraint on heat-engine efficiency,
 conservation laws, and flow topologies gleaned from streamline
 photographs, Jaffer~\cite{thermo3010010} mathematically derives
 convection heat-transfer upper-bound formulas for vertical and
 horizontal flat surfaces.  Actual convective heat-transfer is less
 than these upper bounds when fluid flow is (partially) obstructed.

 Horizontal downward-facing and vertical surfaces are partially
 self-obstructing.  The Jaffer~\cite{thermo3010010} treatment of
 self-obstruction is similar to prior works, but unifies prior work
 self-obstruction factors into a single exact factor which is more
 plausible (measurements not significantly exceeding the upper-bound)
 over the full range of Prandtl numbers (the fluid's momentum
 diffusivity per thermal diffusivity ratio).

\subsection{Characteristic Length}
  The characteristic length $L$ is the length scale of a physical system.
  As with a vertical rectangular plate, a vertical cylinder's
  characteristic length is its height.

 The characteristic length of a level circular cylinder is its
 diameter~$d$.  Generalizing to convex cylinders is the hydraulic
 diameter, which is 4 times the area-to-perimeter ratio of the
 cylinder's cross-section.  Note that the diameter and hydraulic
 diameter are identical for a level circular cylinder.

\subsection{Fluid Mechanics}
  In engineering, convection heat transfer rates are expressed using
  the average surface conductance $\hol$ with units
  ${\rm{W/(m^2\cdot{K})}}$.

  In fluid mechanics, the convective heat transfer rate is represented
  by the dimensionless average Nusselt
  number~($\Nuol\equiv{\hol\,L/k}$), where~$k$ is the fluid's thermal
  conductivity with units ${\rm{W/(m\cdot{K})}}$, and $L$ is the
  system's characteristic length (m).

  The Rayleigh number~$\Ra$ is the impetus for fluid flow due to
  gravity acting on density differences caused by temperature or solute concentration.  A
  fluid's Prandtl number~$\Pra$ is its momentum diffusivity per
  thermal diffusivity ratio.  For mass transfer, the fluid's Schmidt
  number~$\Sc$ is analogous; $\Pra$ will be used in formulas.

  When rising convection induced fluid flow must pass along or around
  the object's surface, $\Ra$ is scaled by a ``self-obstruction''
  factor $1/\Xi$, which depends only on $\Pra$.  This applies to both
  cylinder flow topologies.

  The system's characteristic length~$L$ scales~$\Nuol$, while $L^3$
  scales~$\Ra$.  Variables $\Xi$, $\hol$, $\Pra$, and $\Sc$~are
  independent of~$L$.

\subsection{Flow Topologies}
 Jaffer~\cite{thermo3010010} derived a natural convection formula for
 external flat plates (with convex perimeter) in any orientation from
 its analyses of horizontal and vertical plates.  Similarly, this
 investigation will derive its formula for an inclined cylinder from
 its analyses of horizontal and vertical cylinders.

  There are two topologies of convective flow from external, convex
  cylinders.
  \figref{fig:vertical-flow} shows the induced fluid flows around
  heated vertical and horizontal cylindrical surfaces.

  An important aspect of both flow topologies is that fluid is
  pulled horizontally before being heated by the cylinder.  Pulling
  horizontally expends less energy than pulling vertically because
  the latter does work against the gravitational force.
  Inadequate horizontal clearance around a cylinder can obstruct flow
  and reduce convective heat transfer.

  There is a symmetry in external natural convection; a cooled
  cylinder induces downward flow instead of upward flow.  The rest of
  the present work assumes that the cylindrical exterior surface is
  warmer than the fluid.

\subsection{Turbulence}
  Jaffer~\cite{thermo3010010} derives the formula for an external flat
  surface's total natural convective heat transfer from the
  thermodynamic constraints on heat-engine efficiency, conservation
  laws, and flow topology.

 Thermodynamic and conservation laws make no distinction between
 laminar and turbulent flows.  What about the fluid flow topology?
 If the transition to turbulence is far from a heated convex object,
 then it will not affect heat transfer from the object.  Otherwise,
 the fluid near the object will form a turbulent boundary layer.  This
 turbulent boundary layer will have a viscous sublayer between it and
 the object.
 Regarding forced flow induced boundary layers, Lienhard and
 Lienhard~\cite{ahtt6e}(p.~321) states:

{\narrower
\noindent
 ``Because turbulent mixing is ineffective in the sublayer, the
 sublayer is responsible for a major fraction of the thermal
 resistance of a turbulent boundary layer.''
\par}

\medskip
 The boundary layer thickness grows with decreasing forced velocity.
 Without forced flow, the transition to turbulence will be far
 from the object, increasing turbulent mixing in the boundary layer
 and reducing its thermal resistance.  This makes the non-turbulent
 viscous sublayer responsible for most of the thermal resistance.  If
 the distinction between laminar and turbulent natural convection
 effects heat transfer, its effect will be small.

\medskip
 Prior plate investigations
 assumed that natural convection heat transfer formulas would differ
 substantially when the convection was turbulent versus laminar.  For
 their upward-facing plate, Lloyd and Moran~\cite{lloyd1974natural}
 reported that the transition from laminar to turbulent flow occurred
 at~$\Ra\approx8\times10^6$.  The straight line segments they fitted
 to their data at greater and lesser~$\Ra$ were disjoint at
 $\Ra=8\times10^6$.  However, with their fit lines removed, if
 $\Ra\approx8\times10^6$ represents a discontinuity, then it is one of
 several, and subsumed within the scatter of their measurements (their
 data is plotted in Jaffer~\cite{thermo3010010}).

\medskip
\noindent
  About their measurements of vertical and downward tilted plates,
  Fujii and Imura~\cite{fujii1972natural} wrote:

{\narrower
\noindent
      ``Though the boundary layer was not always laminar near the
      trailing edge for large $\Gr\,\Pra$ [$=\Ra$] values, no influence of the
      flow regime on the data shown in [their] Fig.~6 is appreciable.''
\par}

\medskip
\noindent
  Churchill and Chu~\cite{churchill1975correlating} concludes that one
  of its equations

{\narrower
  \noindent``\dots
  based on the model of Churchill and Usagi~\cite{AIC:AIC690180606}
  provides a good representation for the mean heat transfer for free
  convection from an isothermal vertical plate over a complete range
  of Ra and Pr from 0 to $\infty$ even though it fails to indicate a
  discrete transition from laminar to turbulent flow.''
\par}
\medskip

 Although there are substantial differences between laminar and turbulent
 fluid flow along plates, a single formula appears to govern both
 laminar and turbulent natural convective heat transfer in all
 orientations.

 The flow topology from a vertical cylinder is quite similar to that
 from a vertical plate.  The laminar--turbulent differences in flow
 topology from a level cylinder happen far from the cylinder.  Hence,
 this investigation expects a single formula to govern both laminar
 and turbulent natural convection from cylinders.

\subsection{Combining Transfer Processes}
  Formula~\eqref{eq:mixing} is an unnamed form for combining functions
  which appears frequently in heat or mass transfer formulas:
$$F^p=F_1^p+F_2^p\eqdef{eq:mixing}$$

  Churchill and Usagi~\cite{AIC:AIC690180606} stated that such
  formulas are ``remarkably successful in correlating rates of
  transfer for processes which vary uniformly between these limiting
  cases.''
  Convection transfers heat (or solute) between the cylinder and the fluid.

\subsection{The $\ell^p$-norm}
  When $F_1\ge0$ and $F_2\ge0$, taking the~$p$th root of both sides of
  Equation~\eqref{eq:mixing} yields a vector-space functional form
  known as the
  $\ell^p$-norm, which is notated $\|F_1~,~F_2\|_p$~:
$$\left\|F_1~,~F_2\right\|_p\equiv\left[~|F_1|^p+|F_2|^p\right]^{1/p}\eqdef{eq:l^p}$$

  Norms generalize the notion of distance.  Formally, a vector-space
  norm obeys the triangle inequality: $\|F_1,F_2\|_p\le|F_1|+|F_2|$, which
  holds only for $p\ge1$.  However, $p<1$ is also useful.
 
 When $p>1$, the processes modeled by $F_1$ and $F_2$ compete and
 $\|F_1,F_2\|_p\ge\max(|F_1|,|F_2|)$; the most competitive case is
 $\|F_1,F_2\|_{+\infty}\equiv\max(|F_1|,|F_2|)$.

 The $\ell^1$-norm models independent processes;
 $\|F_1,F_2\|_1\equiv{|F_1|+|F_2|}$.

 When $0<p<1$, the processes cooperate and
 $\|F_1,F_2\|_p\ge{|F_1|+|F_2|}$.  Cooperation between conduction and
 flow-induced heat transfer can occur in natural convection systems.

\section{Data-Sets and Evaluation}

 Heat transfer measurements were captured from graphs in the cited
 works by measuring the distance from each point to its graph's axes,
 then scaling to the graph's units using the ``Engauge'' software
 (version 12.1).

 Churchill and Chu~\cite{CHURCHILL19751049} collected level cylinder
 (angle $\vartheta=0^\circ$) heat and mass transfer measurements from
 eleven studies spanning more than 23 orders of magnitude of~$\Ra$.
 The
 Kutateladze~\cite{KUTATELADZE1963} data-set (with the largest~$\Ra$
 values) is treated separately in \tabref{tab:sources-natural}.

 Nakai Seiichi and Okazaki Takuro~\cite{SEIICHI1975387} measured
 natural convection heat transfer from long horizontal wires with
 $10\times10^3<H/d<12.5\times10^3$ at $\Ra<2\times10^{-5}$.

 Al-Arabi and Khamis~\cite{ALARABI19823} measured natural convection
 heat transfer from a cylinder at six angles.  They measured local
 temperatures along the cylinder, but incorrectly inferred the average
 heat transfer.

 Popiel, Wojtkowiak, and Bober~\cite{POPIEL2007607} measured natural
 convection heat transfer from four vertical cylinders with
 $1<H/d<59$.  Unfortunately, they tested with the bottom of the
 cylinder resting directly on a flat platform, which would impede
 horizontal fluid flow at the bottom.  Extra unheated walls were found
 to significantly affect convective heat transfer in
 Jaffer~\cite{thermo3010010}.

 Goldstein, Khan, and Srinivasan~\cite{GOLDSTEIN2007741} measured
 natural convection mass transfer from three cylinders at four
 inclinations.

 Heo and Chung~\cite{HEO2012366} measured natural convection mass
 transfer from five cylinders with $3.7<H/d\le25$ at inclinations
 $0^\circ\le\vartheta\le90^\circ$.

\medskip
\centerline{\bf\tabdef{tab:sources-natural}\quad{Cylinder natural convection data-sets}}
\vbox{\settabs 8\columns
\toprule
\+\bf Source &&\bf Study&&\hfil $\vartheta$ &~~$\Ra_d/\Xi_\bullet\ge$ &~~$\Ra_d/\Xi_\bullet\le$ & $~~~\,\pm$\hfill\#~&\cr
\midrule
\+Churchill \& Chu~\cite{CHURCHILL19751049} && Kutateladze \cite{KUTATELADZE1963} &&\hfil $ 0^\circ$& ~~$4.3\times10^{9}$& ~~$5.3\times10^{12}$&~~ \hfill  6~&\cr
\+Churchill \& Chu~\cite{CHURCHILL19751049} &&                  all &&\hfil $ 0^\circ$& ~~$7.5\times10^{-12}$& ~~$5.3\times10^{12}$&~~ \hfill 63~&\cr
\+Churchill \& Chu~\cite{CHURCHILL19751049} &&            10 others &&\hfil $ 0^\circ$& ~~$7.5\times10^{-12}$& ~~$3.3\times10^{9}$&~~ \hfill 57~&\cr

\midrule
\+\bf Source && \hfil $\Pra$ or $\Sc$ &\hfil $H/d$ &\hfil $\vartheta$ &~~$\Ra/L^3\ge$ &~~$\Ra/L^3\le$ & $~~~\,\pm$\hfill\#~&\cr
\midrule
\+Nakai \& Okazaki \cite{SEIICHI1975387} &&\hfil $Pr=0.72$ & 10000--12500 &\hfil $0^\circ$& ~~$2.5\times10^{-7}$& ~~$1.5\times10^{-5}$&~~ \hfill 23~&\cr

\midrule
\+AlArabi \& Khamis \cite{ALARABI19823} &&\hfil $Pr=0.708$ &\hfil 15.5--104 &\hfil $ 0^\circ$& ~~$4.8\times10^{9}$& ~~$4.8\times10^{9}$&~~ \hfill  7~&\cr
\+AlArabi \& Khamis \cite{ALARABI19823} &&\hfil $Pr=0.708$ &\hfil 15.5--104 &\hfil $30^\circ$& ~~$4.8\times10^{9}$& ~~$4.8\times10^{9}$&~~ \hfill  7~&\cr
\+AlArabi \& Khamis \cite{ALARABI19823} &&\hfil $Pr=0.708$ &\hfil 15.5--104 &\hfil $45^\circ$& ~~$4.8\times10^{9}$& ~~$4.8\times10^{9}$&~~ \hfill  7~&\cr
\+AlArabi \& Khamis \cite{ALARABI19823} &&\hfil $Pr=0.708$ &\hfil 15.5--104 &\hfil $60^\circ$& ~~$4.8\times10^{9}$& ~~$4.8\times10^{9}$&~~ \hfill  7~&\cr
\+AlArabi \& Khamis \cite{ALARABI19823} &&\hfil $Pr=0.708$ &\hfil 15.5--104 &\hfil $75^\circ$& ~~$4.8\times10^{9}$& ~~$4.8\times10^{9}$&~~ \hfill  7~&\cr
\+AlArabi \& Khamis \cite{ALARABI19823} &&\hfil $Pr=0.708$ &\hfil 15.5--104 &\hfil $90^\circ$& ~~$4.8\times10^{9}$& ~~$4.8\times10^{9}$&~~ \hfill  7~&\cr

\midrule
\+Goldstein et al.~\cite{GOLDSTEIN2007741} &&\hfil $Sc=2300$ &\hfil 0.63--2.34 &\hfil $ 0^\circ$& ~~$3.2\times10^{13}$& ~~$1.6\times10^{14}$&~~ \hfill  4~&\cr
\+Goldstein et al.~\cite{GOLDSTEIN2007741} &&\hfil $Sc=2300$ &\hfil 0.63--2.34 &\hfil $30^\circ$& ~~$1.4\times10^{13}$& ~~$3.2\times10^{14}$&~~ \hfill 11~&\cr
\+Goldstein et al.~\cite{GOLDSTEIN2007741} &&\hfil $Sc=2300$ &\hfil 0.63--2.34 &\hfil $60^\circ$& ~~$6.8\times10^{12}$& ~~$4.4\times10^{14}$&~~ \hfill 11~&\cr
\+Goldstein et al.~\cite{GOLDSTEIN2007741} &&\hfil $Sc=2300$ &\hfil 0.63--2.34 &\hfil $90^\circ$& ~~$2.3\times10^{12}$& ~~$6.1\times10^{14}$&~~ \hfill 16~&\cr

\midrule
\+Heo \& Chung \cite{HEO2012366} &&\hfil $Sc=2094$ &\hfil $25$ &\hfil $0^\circ$--$90^\circ$& ~~$1.3\times10^{14}$& ~~$1.3\times10^{14}$&~~0.9\% \hfill 13~&\cr
\+Heo \& Chung \cite{HEO2012366} &&\hfil $Sc=2094$ &\hfil $7.4$ &\hfil $0^\circ$--$90^\circ$& ~~$1.7\times10^{14}$& ~~$1.7\times10^{14}$&~~0.9\% \hfill 13~&\cr
\+Heo \& Chung \cite{HEO2012366} &&\hfil $Sc=2094$ &\hfil $3.7$ &\hfil $0^\circ$--$90^\circ$& ~~$1.7\times10^{14}$& ~~$1.7\times10^{14}$&~~0.9\% \hfill 13~&\cr
\+Heo \& Chung \cite{HEO2012366} &&\hfil $Sc=2094$ &\hfil $13$ &\hfil $0^\circ$--$90^\circ$& ~~$1.7\times10^{14}$& ~~$1.7\times10^{14}$&~~0.9\% \hfill  9~&\cr
\+Heo \& Chung \cite{HEO2012366} &&\hfil $Sc=2094$ &\hfil $6.7$ &\hfil $0^\circ$--$90^\circ$& ~~$1.7\times10^{14}$& ~~$1.7\times10^{14}$&~~0.9\% \hfill  9~&\cr

\bottomrule
}
\medskip

 These data files, used for generating the present work graphs and
 tables, along with digitization details and estimated digitization
 inaccuracies are available at

 \centerline{\tt https://people.csail.mit.edu/jaffer/convect/NCHTIC-data.zip}.

\subsection{Not Empirical}
  Empirical theories derive their coefficients from measurements,
  inheriting the uncertainties from those measurements.  Theories
  developed from first principles derive exact coefficients and exponents
  mathematically.  For example, Incropera, DeWitt, Bergman, and
  Lavine~\cite{bergman2007fundamentals} (p.~210) gives the thermal
  conductance (units ${\rm{W/K}}$) of a diameter $d$ sphere
  ($L_s=d/2$) into an unbounded stationary, uniform medium having
  thermal conductivity $k$~as:
$$U_0=2\,\pi\,d\,k\eqdef{eq:U0}$$

  The present theory predicting natural convective heat transfer from
  a round cylindrical surface derives from first principles; it is not
  empirical.  Each formula is tied to aspects of the cylinder geometry
  and orientation, fluid, and flow.

\subsection{RMS Relative Error}
 Root-mean-squared (RMS) relative error (RMSRE) provides an objective,
 quantitative evaluation of theory versus experimental data.  It
 gauges the fit of measurements $g(\Ra_j)$ to function $f(\Ra_j)$,
 giving each of the $n$ samples equal weight in
 Formula~\eqref{eq:RMSRE}.
 Along with presenting RMSRE, charts in the present work split RMSRE
 into the bias and scatter components defined in
 Formula~\eqref{eq:bias}.  The root-sum-squared of bias and
 scatter is RMSRE.

$$\eqalignno{{\rm RMSRE}=&\sqrt{{1\over n}\sum_{j=1}^n\left|{g(\Ra_j)\over f(\Ra_j)}-1\right|^2}&\eqdef{eq:RMSRE}\cr
                            {\rm bias}={1\over n}\sum_{j=1}^n\left\{{g(\Ra_j)\over f(\Ra_j)}-1\right\}&\qquad
  {\rm scatter}=\sqrt{{1\over n}\sum_{j=1}^n\left|{g(\Ra_j)\over f(\Ra_j)}-1-{\rm bias}\right|^2}&
  \eqdef{eq:bias}\cr}$$

\section{Theory from Prior Works}

  Subscripts and variable names are not uniform among prior works;
  they have been renamed consistently for inclusion in the present
  work.  Where possible, the formulas are written using the
  $\ell^p$-norm.

\subsection{Vertical Cylinder}
 Sparrow and Gregg~\cite{SPARROW1956} and Cebeci~\cite{CEBECI1974}
 created differential equations modeling the thermal boundary layer
 surrounding a vertical cylinder.  Solved numerically, they created
 graphs and tables relating the cylinder's local temperature profiles
 to a that of a vertical plate having the same height.  Sparrow and
 Gregg's graphs only address $\Pra=0.71$ and $\Pra=1$; Cebeci expanded
 the range to $0.01<\Pra<100$.

 Popiel \etal\cite{POPIEL2007607} fits a complicated
 formula to numerical solutions of the Cebeci~\cite{CEBECI1974}
 equations.

 Of greater interest is a simple formula (similar to the
 vertical plate formula from Churchill and
 Chu~\cite{churchill1975correlating}) that they attribute to
 S. M. Yang~\cite{YANG1985}:
$$\left\|0.36\,{H\over{d}},~0.150\,\root3\of{\Ra_H\over\|1,0.492/\Pra\|_{9/16}}\right\|_{1/2}\eqdef{eq:Yang}$$

 $\|1,0.492/\Pra\|_{9/16}$ is the denominator used for a vertical
 plate in Churchill and Chu~\cite{churchill1975correlating}.
 Jaffer~\cite{thermo3010010} finds that
 $\Xi=\|1,0.5/\Pra\|_{\sqrt{1/3}}$ is more accurate.  The
 self-obstruction factor ($1/\Xi$) for a vertical cylinder and vertical plate should be
 the same; this investigation uses $\Xi=\|1,0.5/\Pra\|_{\sqrt{1/3}}$ for
 vertical cylinders.

\subsection{Level Cylinder}
 Formula~\eqref{eq:Nu-level-cylinder-CC} is the $\ell^p$-norm form of
 what Churchill and Chu~\cite{CHURCHILL19751049} propose as the
 natural convective $\Nuol_d$ from a level diameter~$d$ isothermal
 cylinder (excluding end faces).
$$\left\|0.36\,,~0.150\,\root3\of{\Ra_d\,\over\left\|1,~{0.559/\Pra}\,\right\|_{9/16}}\right\|_{1/2}\eqdef{eq:Nu-level-cylinder-CC}$$

 They also propose a laminar flow formula with an exponent of 1/4
 instead of 1/3:
$$0.36+0.518\,\root4\of{\Ra_d\,\over\left\|1,~{0.559/\Pra}\,\right\|_{9/16}}\eqdef{eq:Nu-level-cylinder-CC-lam}$$

 The denominator $\|1,~{0.559/\Pra}\|_{9/16}$ differs from
 $\|1,0.492/\Pra\|_{9/16}$ only in its coefficient, which is within
 1\% of 9/16.  This investigation uses
 $\Xi_\bullet=\|1,\sqrt{1/3}/\Pra\|_{\sqrt{1/3}}$ as denominator for
 level cylinders.

 Nakai and Okazaki~\cite{SEIICHI1975387} give a formula for tiny $\Ra$
 which has 1.3\% RMSRE on their measurements:
$${3\over\Nu}=\ln\left({3.1\,\sqrt{\Pra+9.4}\over\Nu\,\Pra\,\Ra}\right)\qquad\Pra\Ra\le10^{-3}\eqdef{eq:Seiichi}$$

\subsection{Inclined Cylinder}
 Al-Arabi and Khamis~\cite{ALARABI19823} propose empirical power-law
 Formula~\eqref{eq:AlArabi-laminar} to match their laminar heat
 transfer measurements:
$$\eqalign{&\Nuol_H=\left[2.9-2.32\,\cos^{0.8}\vartheta\right]\,\Gr_d^{1/12}\,\Ra_H^n\qquad n={1\over4}+{\cos^{1.2}\vartheta\over12}\cr
  &1.08\times10^4<Gr_d<6.9\times10^5\qquad9.88\times10^7<Ra_H<4\times10^9\cr}\eqdef{eq:AlArabi-laminar}$$
 and Formula~\eqref{eq:AlArabi-turbulent} to match their turbulent
 measurements:
$$\eqalign{&\Nuol_H=\left[0.47+0.11\,\cos^{0.8}\vartheta\right]\,\Gr_d^{-1/12}\,\Ra_H^{1/3}\cr
  &1.08\times10^4<Gr_d<6.9\times10^5\qquad4\times10^9<Ra_H<2.95\times10^{10}\cr}\eqdef{eq:AlArabi-turbulent}$$

 Al-Arabi and Khamis~\cite{ALARABI19823} does not report observing
 turbulence; they claim to infer its occurrence where the local $h$ at
 the cylinder's end is insensitive to the cylinder's length.

 Heo and Chung~\cite{HEO2012366} also fit empirical power-law curves
 to their data:
$$\eqalignno{
  \Nuol_d&=0.3\,\Ra_d^{0.25}\,[1+0.7\,\cos\vartheta]\qquad\qquad\qquad~~\, {\rm Laminar}&\eqdef{eq:Heo-1}\cr
  \Nuol_d&=0.13\,\Ra_d^{0.3}\,[1+0.6\,\cos\vartheta]\qquad\qquad\qquad~~ {\rm Turbulent}&\eqdef{eq:Heo-2}\cr
  \Nuol_H&=0.67\,\Ra_H^{0.25}\,[1+1.44\,\Ra_d^{-0.04}\,\cos\vartheta]\qquad~{\rm Laminar}&\eqdef{eq:Heo-3}\cr
  \Nuol_H&=0.26\,\Ra_H^{0.28}\,[1+1.89\,\Ra_d^{-0.044}\,\cos\vartheta]\qquad {\rm Turbulent}&\eqdef{eq:Heo-4}\cr
}$$
$$\eqalign{1.69\times10^{8~}&<\Ra_d<5.07\times10^{10}\qquad3.7<H/d<25\cr
           2.64\times10^{12}&<\Ra_H<1.54\times10^{13}\qquad~~~\Pra=2094\cr}
  \eqdef{eq:Heo-0}$$

 The only mention of a turbulence threshold in Heo and
 Chung~\cite{HEO2012366} is $\Gr_H\approx10^9$ which they attribute to
 Al-Arabi and Salman~\cite{ALARABI198045}.  By that criterion, most of
 the $d=0.01$~m cylinder would be laminar, and the others mostly
 turbulent.

 Neither study's formulas include a conduction (heat transfer into
 static fluid) term, so they do not apply to small $\Ra$ values.  Each
 study's formulas apply to a single $\Pra$ or $\Sc$ value, while
 horizontal and vertical cylinder formulas scale $\Ra$ by a function
 of $\Pr$ (or $\Sc$).  Although dimensionally consistent, both of
 these studies mix variables having different characteristic lengths,
 $H$ and $d$, making it difficult to conjecture a corresponding
 physical interpretation.

 Although lacking a conduction term, and proposed only for $\Sc=2300$,
 Goldstein \etal\cite{GOLDSTEIN2007741} addresses this
 characteristic-length issue by proposing a formula using an
 angle-dependent characteristic-length~$G$:
$$\eqalign{&\Shol_G=0.712\,\Ra_G^{1/4}\qquad
  G={H\,\sin\vartheta+d\,\cos\vartheta\over(\sin\vartheta+\cos\vartheta)^{7/3}}\qquad
  \Sc=2300\cr
  &0.63<H/d<2.34\qquad
  2.0\times10^9<\Ra_G<9.5\times10^{11}\cr}
  \eqdef{eq:Goldstein}$$
 where $\Shol$ is the (average) Sherwood number, an analog of the
 (average) Nusselt number~$\Nuol$.

 About Formula~\eqref{eq:Goldstein}, Goldstein \etal\cite{GOLDSTEIN2007741}
 writes:

{\narrower
  \noindent
  The correlation suggests that the flow is in the laminar region for
  the most part.  The horizontal cylinder data taken alone suggest the
  onset of turbulent flow; however, within the range of the Rayleigh
  numbers studied, the correlation above will give a Sherwood number
  that is reasonably accurate.
\par}

 The approach of this investigation is to derive the general formulas
 for horizontal and vertical cylinders from thermodynamic constraints
 on heat-engine efficiency, conservation laws, and flow topology.
 Streamline photos of natural convection from a cylinder were not
 found.  Experience developing the plate natural convection flow
 models led to the flows proposed in \figrefs{fig:vertical-flow}{a}
 and \figrefn{fig:vertical-flow}{b}.

 Horizontal and vertical flow modes compete in the case of an inclined
 plate.  This investigation posits that the same is true for an
 inclined cylinder.

\section{Natural Convection}

  First, a review of flat surface convection (which will also be used
  for cylinder end-caps):

  From thermodynamic constraints and conservation laws,
  Jaffer~\cite{thermo3010010} derives generalized natural convection
  Formula~\eqref{eq:general} with the parameters specified
  in \tabref{tab:natural convection parameters}:
$$\Nuol=\left\|\Nuz\,\bigl[1-C\bigr], \root{2+E}\of{\bigl[{C\,D\,\Nuz}\bigr]^{3+E}{{2\over B}\,\Ra}}~\right\|_p\eqdef{eq:general}$$
$$\Xi=\left\|1~,~{1/2\over\Pra}\right\|_{\sqrt{1/3}}\quad\Nuzs={2\over\pi}\approx0.637 \quad\Nuzq={2^4\over\root4\of2\,\pi^2}\approx1.363\eqdef{eq:Xi}$$

 The $*$ superscript indicates an upward-facing surface; the $'$
 superscript is for vertical surfaces; the $R$ subscript is for a
 downward-facing surface.  Scripted $\Nuol$ and $\Ra$ are computed
 with $L$ having the same script.

\medskip
\centerline{\bf\tabdef{tab:natural convection parameters}\quad{Plate natural convection parameters}}
\vbox{\settabs 11\columns
\toprule
  \+\bf Face &\hfill$\theta$~~\qquad&$L$  &$\Nuol$   &$\Nuz$   &$\Ra$     &\hfil$B$    &\hfil$C$       &\hfil$D$  &\hfil$E$&\hfil$p$  &\cr
\midrule                                                                                                          
  \+ up      &\hfill$-90^\circ$\qquad&$\Ls$ &$\Nuols$ &$\Nuzs$  &$\Ra^*$    &\hfil$2$    &\hfil$1/\sqrt8$&\hfil$1$  &\hfil1  &\hfil$1/2$&&\cr
  \+ vertical&\hfill  $0^\circ$\qquad&$L'$&$\Nuolq$ &$\Nuzq$  &$\Ra'/\Xi$ &\hfil$1/2$  &\hfil$1/2$     &\hfil$1/4$&\hfil1  &\hfil$1/2$&&\cr
  \+ down    &\hfill$+90^\circ$\qquad&$L_R$ &$\NuolR$ &$\Nuzq/2$&$\Ra_R/\Xi$&\hfil$4$    &\hfil$1/2$     &\hfil$2$  &\hfil3  &\hfil$1$  &&\cr
\bottomrule
}

The parameters of \tabref{tab:natural convection parameters} are:

\unorderedlist

 \li $\theta$ is the angle of the plate from vertical;
 $\theta=+90^\circ$ is face down.

 \li $L$ is the characteristic length of a flat plate with convex
 perimeter:

\unorderedlist

 \li face up ($\theta=-90^\circ$) $\Ls$ is the area-to-perimeter ratio;

\unorderedlist
  \li For a $Y\times Z$ rectangle, $\Ls=Y\,Z/[2\,Y+2\,Z]$.
  \li For a diameter $d$ disk, $\Ls=d/4$.
\endunorderedlist

 \li vertical ($\theta=0^\circ$) $L'$ is the harmonic mean of the
 perimeter vertical spans;

\unorderedlist
  \li Given a convex vertical plate with maximum width $W$ whose
  vertical span at horizontal offset $w$ is $H(w)$:
$$L'=W\left/\int_0^W{1\over H(w)}\,\diff{w}\right.\eqdef{eq:L'}$$
  \li For a diameter $d$ disk, $L'=2\,d/\pi$.
  \li For a level height $H$ rectangle, $L'=H$.
\endunorderedlist

 \li face down ($\theta=+90^\circ$) $L_R$ is the harmonic mean of the
 perimeter distances to that bisector which is perpendicular to the
 shortest bisector;

\unorderedlist
  \li Given a flat surface with convex perimeter defined by functions
  $H_+(w)>0$ and $H_-(w)<0$ within the range $0<w<W$ along the
  equal-area bisector which is perpendicular to the shortest
  equal-area bisector:
$$L_R=4\,W\left/\int_0^W\left[{1\over |H_+(w)|}+{1\over |H_-(w)|}\right]\,\diff{w}\right.\eqdef{eq:L_R}$$
  \li For a $Y\times Z$ rectangle, $L_R=\min(Y,Z)/2$.
  \li For a diameter $d$ disk, $L_R=d/\pi$.
\endunorderedlist

\endunorderedlist

 \li $\Nuol$ is the dimensionless convective heat transfer
   from Formula~\eqref{eq:general}.

 \li $\Nuz$ is the dimensionless heat conduction into motionless fluid.

 \li $\Ra$ Rayleigh number: $\Ra'$ is computed with vertical $L'$;
  $\Ra^*=\Ra'\,[\Ls/L']^3$; $\Ra_R=\Ra'\,[L_R/L']^3$.

  $\Pra$ does not affect upward-facing heat transfer because the
  heated fluid flows directly upward, as does conducted heat.  When
  heated fluid must flow along vertical and downward-facing plates,
  its heat transfer potential is reduced by dividing $\Ra$ by $\Xi$
  from Formula~\eqref{eq:Xi}.

 \li $B$~is the sum of the mean lengths of flows tangent to a wall or
  counter-flow on one side divided by~$L$.

 \li $C$~is the surface area fraction responsible for flow induced
  heat transfer.

 \li $D$~is the effective length of heat transfer contact with one
  side of the plate divided by~$L$.

 \li $E$~is the count of $90^\circ$ changes in direction of thermally
  induced fluid flow.

 \li $p$: The $\ell^p$-norm combines the static conduction and induced
  convective heat flows.

\endunorderedlist

\subsection{Cylinder}
 Formula~\eqref{eq:general} is not limited to flat surfaces.  Natural
 convection heat transfer from a cylinder can be modeled using
 Formula~\eqref{eq:general} with the parameters
 in \tabref{tab:cylinder convection parameters}.

 The $\bullet$ superscript or subscript indicates a level cylinder;
 the $\|$ superscript is for a vertical cylinder.

\medskip
\centerline{\bf\tabdef{tab:cylinder convection parameters}\quad{Cylinder natural convection parameters}}
\vbox{\settabs 11\columns
\toprule
  \+\bf Cylinder &\hfill$\vartheta$~\qquad&$L$  &$\Nuol$   &$\Nuz$   &$\Ra$    &\hfil$B$              &\hfil$C$   &\hfil$D$  &\hfil$E$&\hfil$p$  &\cr
\midrule                                                                                                                   
 \+ level   & \hfill$0^\circ$\qquad&$d$&$\Nuolb$&$\Nuzb$           &$\Ra_d/\Xi_\bullet$&\hfil$\pi/2$&\hfil$1/2$&\hfil$\root{8}\of{2}$ &\hfil$E^\bullet$&\hfil$1/3$&\cr
 \+ vertical&\hfill$90^\circ$\qquad&$H$&$\Nuolp$&$\Nuzb{H\over{d}}$&$\Ra_H/\Xi$&\hfil${H\over{d}}$&\hfil$1/2$&\hfil${\root{8}\of{2}\over2\,\pi}{d\over{H}}$&\hfil$1$&\hfil$1/6$&\cr
\bottomrule
}

The parameters of \tabref{tab:cylinder convection parameters} are:

\unorderedlist

 \li $\vartheta$ is the angle of the cylinder from horizontal.

 \li $L$: The height $H$ of a vertical cylinder is its characteristic
 length.

 The diameter~$d$ of a level circular cylinder is its characteristic
 length.  This investigation uses the hydraulic diameter, which is 4
 times the area-to-perimeter ratio of the cylinder's cross-section,
 as~$d$.  Note that the diameter and hydraulic diameter are identical
 for a level circular cylinder.

 \li $\Nuol$ is the dimensionless convective heat transfer from
   Formula~\eqref{eq:general}.

 \li $\Nuz$ is the dimensionless heat conduction into motionless
   fluid.  Conduction shape factors are not well-defined with
   unbounded source areas, but Nusselt numbers can be:
$$\Nuzb={\root{8}\of{2}\over\pi}\approx0.3471\qquad
\left[{1\over\sqrt{8}}\approx0.354 {\rm~in~earlier~versions}\right]\eqdef{eq:Nu0-cylinder}$$
 Note that $\Nuzb/2\approx0.173$ is much smaller than the 0.36 value in
 prior works Formulas~\eqref{eq:Yang}, \eqref{eq:Nu-level-cylinder-CC},
 and \eqref{eq:Nu-level-cylinder-CC-lam}.

 The vertical cylinder's static conduction is the same as a level cylinder.
 But its characteristic length is $H$, not $d$.  Consequently vertical
 $\Nuz=\Nuzb\,H/d$ as per S. M. Yang~\cite{YANG1985} in
 Formula~\eqref{eq:Yang}.

 \li $\Ra$ Rayleigh number: vertical $\Ra_H$ is computed with
 vertical~$H$; level $\Ra_d=\Ra_H\,[d/H]^3$.

 $\Ra_H$ is divided by Formula~\eqref{eq:Xi} $\Xi(Pr)$ as with a
 vertical plate.

 The upper part of the level cylinder is unobstructed; with small
 $\Pra$, induced fluid flow and heat transfer are reduced.  The
 $\Ra_d$ divisor is increased to Formula~\eqref{eq:Xi-cyl}
 $\Xi_\bullet$.  \figref{fig:Xis} plots $1/\Xi_\bullet$ and $1/\Xi$.
$$\Xi_\bullet=\left\|1~,~{\sqrt{1/3}\over\Pra}\right\|_{\sqrt{1/3}}\qquad
  \Xi=\left\|1~,~{1/2\over\Pra}\right\|_{\sqrt{1/3}}\eqdef{eq:Xi-cyl}$$

\vfill\eject

\vbox{\settabs 1\columns
\+\hfil\figscale{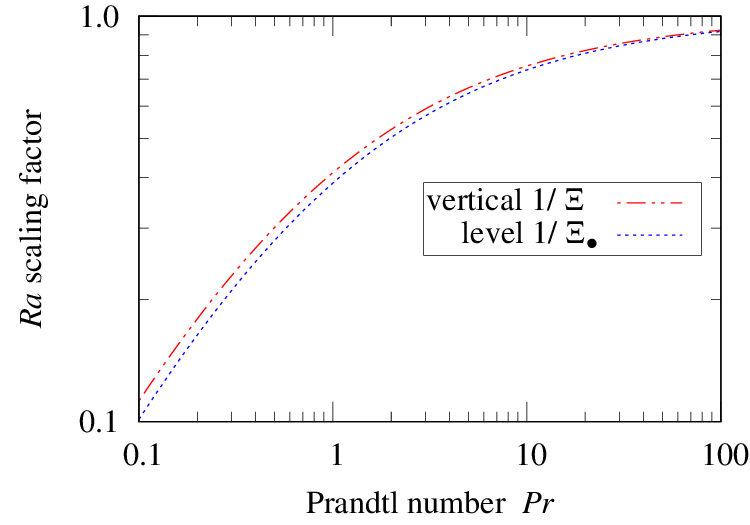}{234pt}&\cr
\+\hfil{\bf\figdef{fig:Xis}\quad{$\Ra$ scaling factors}}&\cr
}
\medskip

 \li $B$: Fluid rises the full height of the vertical cylinder;
 its~$B$ would be expected to be~1.  Like $\Nuz$ however, vertical $B$
 gets scaled by $H/d$.

 Fluid flows tangent to the horizontal cylinder over (each) half of
 its perimeter in \figref{fig:vertical-flow}{b}; $B=\pi/2$.

 \li $C$: $C=1/2$ because fluid flow with both cylinder orientations is
 two-dimensional.

 \li $D$:
 Most of the level cylinder's heat transfer occurs where its surface
 is downward-facing:
$$D=\root{8}\of{2}\approx1.091\qquad
  \left[{\pi\over3}\approx1.047 {\rm~in~earlier~versions}\right]
  \eqdef{eq:level-D}$$

 The vertical plate's $D=1/4$ because fluid flowing by the upper half
 is already heated and accelerating upward.  The vertical cylinder
 lacks vertical edges adjacent to unheated fluid; Relative to $H$,
 only a short length of the vertical cylinder heats a significant
 amount of fluid:
$$D={\root{8}\of{2}\over2\,\pi}\,{d\over{H}}\qquad{\root{8}\of{2}\over2\,\pi}\approx0.1736\qquad
  \left[{1\over6}\approx0.1667 {\rm~in~earlier~versions}\right]
  \eqdef{eq:vertical-D}$$
 The ${d/{H}}$ factor in Formula~\eqref{eq:vertical-D} unscales
 $\Nuz=\Nuzb\,H/d$ in the $C\,D\,\Nuz$ term of
 Formula~\eqref{eq:general}.

 \li $E$: Fluid flow induced by a vertical cylinder experiences one
 $90^\circ$ ($\pi/2$~rad) change in direction; its~$E=1$.

 \figref{fig:vertical-flow}{b} shows convecting fluid flowing tangent
 to the lower half of a level cylinder.  For a unit radius cylinder,
 fluid flowing horizontally at elevation $0<y<1$ must bend
 $\pi-\arcsin({y})$ radians upward in order to be tangent to the
 cylinder.  The average angle is:
$$\int_0^1\pi-\arcsin{y}\,\diff{y}
  =1+{\pi\over2}\approx2.571\rm~rad\eqdef{eq:average-bottom-angle}$$

 Fluid at mid-height is already moving upward; it requires no more
 bend.  Fluid moving tangent to the upper part of a cylinder will
 require some bend, but less fluid is flowing tangent to the cylinder
 at the top; so $\arcsin{x}$ is weighted by $x^2$, where $x$ is the
 distance from the vertical mid-line:
$$\int_0^1{\pi\over2}-x^2\,\arcsin{x}\,\diff{x}
  ={2+3\,\pi\over9}\approx1.269\rm~rad\eqdef{eq:average-top-angle}$$

 Dividing the average of Formula~\eqref{eq:average-bottom-angle} and
 Formula\eqref{eq:average-top-angle} by $\pi/2$:
$$E^\bullet={1\over2}+{1\over\pi}+{1\over3}+{2\over9\,\pi}
          ={5\over6}+{11\over9\,\pi}\approx1.222\eqdef{eq:E-bullet}$$

 \li $p$: Lacking vertical edges adjacent to unheated fluid, the
 static conduction and induced convective heat flows of a vertical
 cylinder cooperate much more than from a vertical plate; they combine
 as the $\ell^{1/6}$-norm.

 The downward-facing part of a level cylinder preheats fluid which
 then flows adjacent to the vertical surfaces.
 The level cylinder's heat flows cooperate more than the vertical
 plate's $\ell^{1/2}$-norm, but less than the vertical cylinder's
 $\ell^{1/6}$-norm.  They combine as the $\ell^{1/3}$-norm.

\endunorderedlist

\vfill\eject

\subsection{Surface Conductance Formulas}
 Because the characteristic lengths scaling a flat plate's $\Nuols$,
 $\Nuolq$, and $\NuolR$ can be different, flow-mode interactions use
 scale-free $\hsol$, $\hqol$, and $\hRol$:
$$\hsol={k\over\Ls}\,\left\|\Nuzs\,\left[1-\sqrt{1\over{8}}\right]~,~{\Nuzs^{4/3}\over4}\,\root3\of{\Ra^*}\right\|_{1/2}
  \approx{k\over\Ls}\,\left\|0.411~,~0.137\,\root3\of{\Ra^*}\right\|_{1/2}\eqdef{eq:upward}$$
$$\hqol={k\over{L'}}\,\left\|{{\Nuzq}\over2}~,~{\Nuzq^{4/3}\over8\,\root3\of2}\,\root3\of{\Ra'\over\Xi}\right\|_{1/2}
  \approx{k\over{L'}}\,\left\|0.682~,~0.150\,\root3\of{\Ra'\over\Xi}\right\|_{1/2}\eqdef{eq:vertical}$$
$$\hRol={k\over{L_R}}\,\left\|{\Nuzq\over4}~,~{\Nuzq^{6/5}\over2^{7/5}}\,\root5\of{\Ra_R\over\Xi}\right\|_1
  \approx{k\over{L_R}}\,\left[0.341+0.550\,\root5\of{\Ra_R\over\Xi}\right]\eqdef{eq:downward}$$

 Formulas~\eqref{eq:level-cylinder} and~\eqref{eq:vertical-cylinder}
 model heat transfer from horizontal and vertical cylinders, respectively.
$$\holb={k\over{d}}\,\left\|{\Nuzb\over2}~,~\root{2+E^\bullet}\of{\left[{\root{8}\of{2}\,\Nuzb\over2}\right]^{3+E^\bullet}{\Ra_d\over\pi\,\Xi_\bullet}}\right\|_{1/3}
  \approx{k}\,\left\|{0.174\over{d}}~,~{0.0792\over{d}}\,\left[{\Ra_d\over\Xi_\bullet}\right]^{0.310}\right\|_{1/3}\eqdef{eq:level-cylinder}$$
$$\holp={k\over{H}}\,\left\|{\Nuzb\over2}{H\over{d}}~,~\root3\of{\left[{\root{8}\of{2}\,\Nuzb\over4\,\pi}\right]^4{d\over{H}}\,{2\,\Ra_H\over\Xi}}\right\|_{1/6}
  \approx{k}\,\left\|{0.174\over{d}}~,~{0.01181\over{H}}\,\root3\of{{d\over{H}}\,{\Ra_H\over\Xi}}\right\|_{1/6}\eqdef{eq:vertical-cylinder}$$

\subsection{End-Caps}
 The flow modes of each end-cap (upper and lower, respectively)
 compete as the $\ell^{16}$-norm:
$$\huol=\left\|\hqol(\cos\vartheta\,\Ra_H),~{\hsol(\sin\vartheta\,\Ra^*)\over1+H/d}\right\|_{16}\qquad
  \hdol=\left\|\hqol(\cos\vartheta\,\Ra_H),~{\hRol(\sin\vartheta\,\Ra_R)\over1+H/d}\right\|_{16}
  \eqdef{eq:end-caps}$$
 where $L'=2\,d/\pi$, $\Ls=d/4$, and $L_R=d/\pi$.  The $1+H/d$
 denominator models the reduction of vertical cylinder heat transfer
 by already heated fluid.
 The trigonometric functions of $\vartheta$ are explained
 in \ref{Inclination}.
 Note that the end caps and cylinder surface have different areas.

\subsection{Comparison With Level Cylinder Measurements}
 \figref{fig:natural-level} compares five level cylinder theories with
 data-sets from Churchill and Chu~\cite{CHURCHILL19751049},
 Goldstein \etal\cite{GOLDSTEIN2007741}, Heo and
 Chung~\cite{HEO2012366}, and Nakai and Okazaki~\cite{SEIICHI1975387}.
 Note that the estimated digitization accuracy of the Churchill and
 Chu~\cite{CHURCHILL19751049} data is $\pm10\%$.

\unorderedlist
 \li The ``Churchill \& Chu: turbulent (theory)'' trace is
 Formula~\eqref{eq:Nu-level-cylinder-CC}.

 \li The ``Churchill \& Chu: laminar (theory)'' trace is
 Formula~\eqref{eq:Nu-level-cylinder-CC-lam}.

 \li The ``Nakai and Okazaki 1975 (theory)'' trace is
 Formula~\eqref{eq:Seiichi}.

 \li The ``Goldstein et al.~2007 (theory)'' trace is
 Formula~\eqref{eq:Goldstein}.

 \li The ``present work'' trace is $\Nuolb=\holb\,d/k$, where $\holb$
 is Formula~\eqref{eq:level-cylinder}.
\endunorderedlist

 \tabref{tab:CC-conformance} presents statistics of the Churchill and
 Chu data-set measurements versus turbulent
 Formula~\eqref{eq:Nu-level-cylinder-CC}, laminar
 Formula~\eqref{eq:Nu-level-cylinder-CC-lam}, and the present work's
 Formula~\eqref{eq:level-cylinder}.

\medskip
\centerline{\bf\tabdef{tab:CC-conformance}\quad{Churchill and Chu data-sets versus theories}}
\vbox{\settabs 9\columns
\toprule
\+\bf Study &&\bf Theory&$\Ra_d/\Xi_\bullet\ge$ &~~$\Ra_d/\Xi_\bullet\le$&\hfill\bf RMSRE&\hfill\bf Bias &\hfill\bf Scatter&\hfill\bf\#~&\cr
\midrule
\+Kutateladze \cite{KUTATELADZE1963} && turbulent& $4.3\times10^{9}$ &~~$5.3\times10^{12}$&\hfill $ 7.4\%$ &\hfill $ -1.4\%$ &\hfill $ 7.3\%$ &\hfill  6~&\cr
\+Kutateladze \cite{KUTATELADZE1963} && laminar& $4.3\times10^{9}$ &~~$5.3\times10^{12}$&\hfill $185.6\%$ &\hfill $-172.9\%$ &\hfill $67.4\%$ &\hfill  6~&\cr
\+Kutateladze \cite{KUTATELADZE1963} && present& $4.3\times10^{9}$ &~~$5.3\times10^{12}$&\hfill $94.7\%$ &\hfill $-90.0\%$ &\hfill $29.4\%$ &\hfill  6~&\cr
\+                 all && turbulent& $7.5\times10^{-12}$ &~~$5.3\times10^{12}$&\hfill $20.2\%$ &\hfill $ +8.8\%$ &\hfill $18.2\%$ &\hfill 63~&\cr
\+                 all && laminar& $7.5\times10^{-12}$ &~~$5.3\times10^{12}$&\hfill $60.1\%$ &\hfill $-23.7\%$ &\hfill $55.2\%$ &\hfill 63~&\cr
\+                 all && present& $7.5\times10^{-12}$ &~~$5.3\times10^{12}$&\hfill $31.2\%$ &\hfill $ -9.5\%$ &\hfill $29.7\%$ &\hfill 63~&\cr
\+           10 others && turbulent& $7.5\times10^{-12}$ &~~$3.3\times10^{9}$&\hfill $21.1\%$ &\hfill $ +9.8\%$ &\hfill $18.7\%$ &\hfill 57~&\cr
\+           10 others && laminar& $7.5\times10^{-12}$ &~~$3.3\times10^{9}$&\hfill $19.0\%$ &\hfill $ -7.9\%$ &\hfill $17.2\%$ &\hfill 57~&\cr
\+           10 others && present& $7.5\times10^{-12}$ &~~$3.3\times10^{9}$&\hfill $11.4\%$ &\hfill $ -1.0\%$ &\hfill $11.3\%$ &\hfill 57~&\cr

\bottomrule
}
\medskip

\vbox{\settabs 1\columns
\+\hfil\figscale{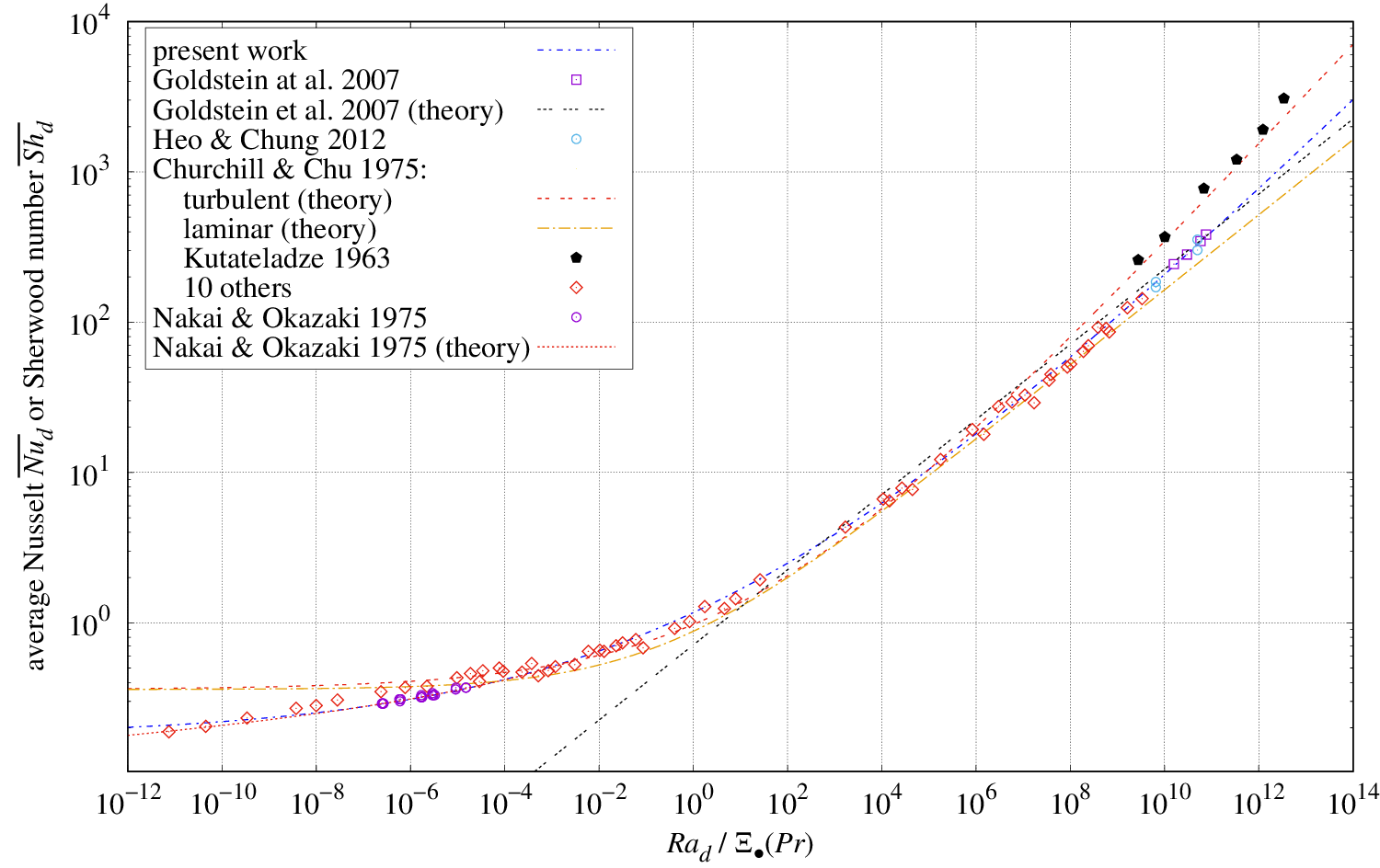}{468pt}&\cr
\+\hfil{\bf\figdef{fig:natural-level}\quad{Natural convection from level cylinder}}&\cr
}
\medskip

\vbox{\settabs 1\columns
\+\hfil\figscale{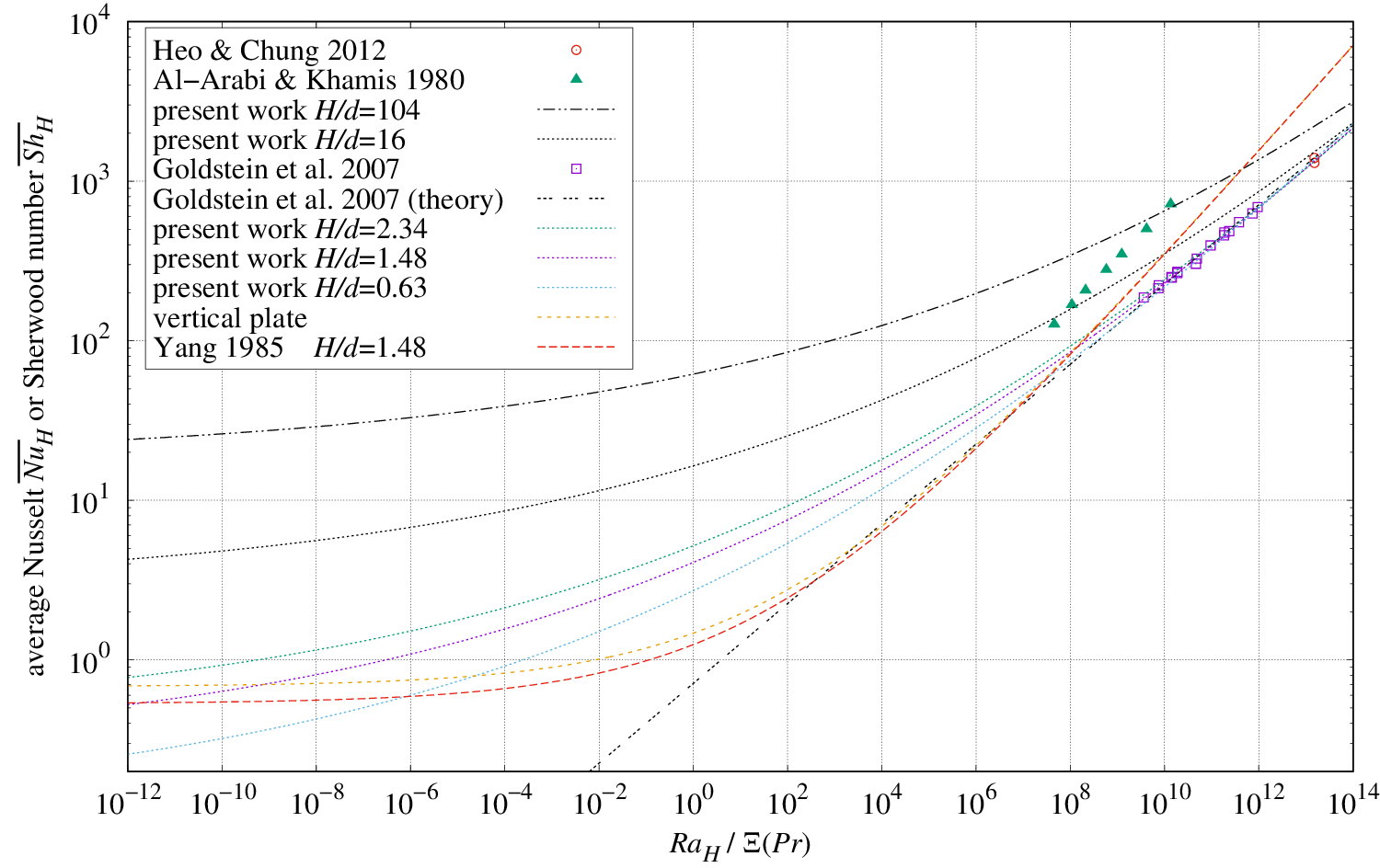}{468pt}&\cr
\+\hfil{\bf\figdef{fig:natural-vertical}\quad{Natural convection from vertical cylinder}}&\cr
}
\vfill\eject

 The six point data-set which Churchill and
 Chu~\cite{CHURCHILL19751049} attribute to
 Kutateladze~\cite{KUTATELADZE1963} exceeds all the theory traces.
 The cylinder convection graph in Kutateladze~\cite{KUTATELADZE1963}
 does not assign parameters or references to individual measurements,
 but asserts that $\Pra\ge1$.  The large $\Ra_d$ measurements appear
 to be from Hermann~\cite{Hermann1936}, which was ``accepted as
 dissertation by the Technical High School at Aachen'' in 1936.  The
 two studies with large $\Ra_d$ which Hermann cites are
 Wamsler~\cite{Wamsler1911} and Koch and Beih~\cite{Koch1927}.
 Hermann gives $\Pra=0.74$; however, this $\Pra$ difference is not
 sufficient to bring the measurements into line with the other
 Churchill and Chu data-sets.

 Both the Wamsler~\cite{Wamsler1911} and Koch and Beih~\cite{Koch1927}
 data-sets were obtained from heated iron and steel pipes in air.
 Hermann states that the radiative heat loss ranged up to 60\% of the total,
 and that the radiative losses of Koch and Beih had to be corrected.
 Having 60\% of the heat loss due to thermal radiation degrades the
 potential accuracy of convection measurements.

 Solute-transfer measurements of natural convection are capable of
 greater accuracy.  Heo and Chung~\cite{HEO2012366} estimates $0.9\%$
 measurement uncertainty.  In \tabref{tab:level-conformance} the
 Goldstein \etal\cite{GOLDSTEIN2007741} and Heo and
 Chung~\cite{HEO2012366} data sets have RMSRE less than~5\%.
 The $\Ra/\Xi_\bullet$ range of the Goldstein \etal and Heo and Chung
 data-sets overlap three of the six Kutateladze~\cite{KUTATELADZE1963}
 points, providing further evidence justifying the exclusion of the
 Kutateladze data.

 The remaining ten of the Churchill and Chu data-sets,
 spanning more than 20 orders-of-magnitude of $\Ra/\Xi_\bullet$, have
 11\% RMSRE versus the present theory in \tabref{tab:CC-conformance},
 a significant improvement from the 20\% RMSRE of
 Formula~\eqref{eq:Nu-level-cylinder-CC} and the 19\% RMSRE of
 Formula~\eqref{eq:Nu-level-cylinder-CC-lam}.

 The ``Goldstein et al.~(theory)'' trace matches its own level
 cylinder data with 3.6\% RMSRE; but, lacking a heat conduction term,
 would be inaccurate when $\Ra/\Xi_\bullet<10$.

\medskip
\centerline{\bf\tabdef{tab:level-conformance}\quad{Level cylinder measurements versus present theory}}
\vbox{\settabs 9\columns
\toprule
\+\bf Source && \hfil$\Pra$ or $\Sc$&\hfil $H/d$ &\hfil $\vartheta$ &\hfill\bf RMSRE&\hfill\bf Bias &\hfill\bf Scatter&\hfill\bf\#~&\cr
\midrule
\+Goldstein et al.~\cite{GOLDSTEIN2007741} &&\hfil $Sc=2300$ &\hfil 0.63--2.34 &\hfil $ 0^\circ$&\hfill $ 2.4\%$ &\hfill $ -1.8\%$ &\hfill $ 1.6\%$ &\hfill  4~&\cr

\+Heo \& Chung \cite{HEO2012366} &&\hfil $Sc=2094$ &\hfil 3.73--13.2&\hfil $0^\circ$&\hfill $ 2.9\%$ &\hfill $ +2.6\%$ &\hfill $ 1.3\%$ &\hfill  4~&\cr

\+Nakai \& Okazaki \cite{SEIICHI1975387} &&\hfil $Pr=0.72$ & 10000--12500 &\hfil $0^\circ$&\hfill $ 1.5\%$ &\hfill $ -1.1\%$ &\hfill $ 1.0\%$ &\hfill 23~&\cr

\bottomrule
}
\medskip

\subsection{Comparison With Vertical Cylinder Measurements}
 \figref{fig:natural-vertical} compares four vertical cylinder
 theories with data-sets from Al-Arabi and Khamis~\cite{ALARABI19823},
 Goldstein \etal\cite{GOLDSTEIN2007741}, and Heo and
 Chung~\cite{HEO2012366}.

\unorderedlist
 \li The ``vertical plate'' trace is $\Nuolq=\hqol\,L'/k$, where $\hqol$ is
 Formula~\eqref{eq:vertical}.

 \li The ``Yang 1985'' trace is Formula~\eqref{eq:Yang}.

 \li The ``Goldstein et al. (theory)'' trace is
 Formula~\eqref{eq:Goldstein}.

 \li The ``present work'' traces are $\Nuolp=\holp\,H/k$, where
 $\holp$ is Formula~\eqref{eq:vertical-cylinder}.
\endunorderedlist

 \figref{tab:vertical-conformance} presents vertical statistics of the
 inclined cylinder data-sets.  Comparing vertical cylinder heat
 transfer measurements is more difficult than with level cylinders
 because the surface conductance depends on $H/d$, the
 height-to-diameter ratio.

 The ``Goldstein et al.~(theory)'' trace matches its own level
 cylinder data with 3.6\% RMSRE.  However, the cylinder diameter $d$
 does not affect the ``Goldstein et al.~(theory)''
 Formula~\eqref{eq:Goldstein} when the cylinder is vertical.  Its
 RMSRE for vertical Al-Arabi and Khamis data-sets exceeds~100\%.

 While the other formulas all raise $\Ra_H^{1/3}$, the gentle
 curvature of the $\ell^{1/6}$-norm dominates the slopes of ``present
 work'' traces through the entire range
 of \figref{fig:natural-vertical}.  This indicates that vertical plate
 convection is rarely a good approximation for vertical cylinder
 convection.

\medskip
\centerline{\bf\tabdef{tab:vertical-conformance}\quad{Vertical cylinder measurements versus present theory}}
\vbox{\settabs 9\columns
\toprule
\+\bf Source && \hfil$\Pra$ or $\Sc$ &\hfil $H/d$&\hfil $\vartheta$ &\hfill\bf RMSRE&\hfill\bf Bias &\hfill\bf Scatter&\hfill\bf\#~&\cr
\midrule
\+Goldstein et al.~\cite{GOLDSTEIN2007741} &&\hfil $Sc=2300$ &\hfil 0.63--2.34 &\hfil $90^\circ$&\hfill $ 2.8\%$ &\hfill $ -1.1\%$ &\hfill $ 2.6\%$ &\hfill 16~&\cr

\+Heo \& Chung \cite{HEO2012366} &&\hfil $Sc=2094$ &\hfil 3.73--13.2&\hfil $90^\circ$&\hfill $ 2.8\%$ &\hfill $ +1.0\%$ &\hfill $ 2.6\%$ &\hfill  4~&\cr

\+AlArabi \& Khamis \cite{ALARABI19823} &&\hfil $Pr=0.708$ &\hfil 15.5--104 &\hfil $90^\circ$&\hfill $ 4.1\%$ &\hfill $ +0.3\%$ &\hfill $ 4.0\%$ &\hfill  7~&\cr

\bottomrule
}
\medskip

\section{Inclination}

\indent
 $\theta$ is the angle of a flat surface from vertical;
 $\theta=-90^\circ$ is face up.

 $\vartheta$ is the angle of a cylinder's axis from horizontal.

 A mass constrained to move along a line inclined at $\theta$ from
 vertical will experience gravitational force proportional to the
 projection of the gravity vector onto that line, $|\cos\theta|$.
 Similarly, a mass constrained to move perpendicular to a plate
 inclined at $\theta$ from vertical will have its force scaled by
 $|\sin\theta|$.

 $\Ra$ is proportional to gravitational acceleration in the direction
 of flow; thus plate $\Ra'$ is scaled by $|\cos\theta|$ and $\Ra^*$ and
 $\Ra_R$ get scaled by $|\sin\theta|$.
 Similarly for cylinders, $\Ra_H$ gets scaled by $|\sin\vartheta|$ and
 $\Ra_d$ gets scaled by $|\cos\vartheta|$.

 \subsection{Natural Convection From an Inclined Plate}
 For an inclined plate, the formula in Raithby and
 Hollands~\cite{rohsenow1998handbook} chooses the upward-facing,
 downward-facing, or vertical flow mode having the maximum convective
 surface conductance (with each $\Ra$ scaled as described above).

 The upward-facing $\hsol$ and downward-facing $\hRol$ do not directly
 compete with each other, suggesting:
$$\hol=\cases{
 \max\left({\hqol({\left|\cos\theta\right|}\,\Ra')},~{\hsol({\left|\sin\theta\right|}\,\Ra^*)}\right)&$\sin\theta<0$\cr
 \max\left({\hqol({\left|\cos\theta\right|}\,\Ra')},~{\hRol\left({\left|\sin\theta\right|}\,\Ra_R\right)}\right)&$\sin\theta\ge0$\cr}
 \eqdef{eq:inclined-max}$$

 However, measurements of inclined plate natural convective heat
 transfer revealed that, in reality, the~$\theta$ transition is more
 gradual using the $\ell^{16}$-norm in
 Formula~\eqref{eq:inclined-natural}:
$$\hol=\cases{
 \left\|{\hqol({\left|\cos\theta\right|}\,\Ra')},~{\hsol({\left|\sin\theta\right|}\,\Ra^*)}\right\|_{16}&$\sin\theta<0$\cr
 \left\|{\hqol({\left|\cos\theta\right|}\,\Ra')},~{\hRol\left({\left|\sin\theta\right|}\,\Ra_R\right)}\right\|_{16}&$\sin\theta\ge0$\cr}
 \eqdef{eq:inclined-natural}$$

\subsection{Natural Convection From an Inclined Cylinder}
 Flow along a flat surface is strongly constrained by that surface;
 competing flows combine with the $\ell^{16}$-norm ($p=16$).  Flows
 around an inclined cylinder are less constrained but still compete,
 suggesting a smaller~$p>1$.

 However, natural convection flows around a long thin cylinder will be
 more competitive ($p\gg1$) than from a cylinder where~$H\approx{d}$.
 This suggests combining $\holp$ and $\holb$ with $p=1+H/d$:
$$\hol=\left\|{\holp\left({\left|\sin\vartheta\right|}\,{\Ra_H}\right)},~{\holb\left({\left|\cos\vartheta\right|}\,{\Ra_d}\right)}\right\|_{1+H/d}
   \eqdef{eq:inclined-cylinder-natural0}$$

 Formula~\eqref{eq:inclined-cylinder-natural0} is suitable for
 $\Ra/\Xi\gg1$, but doubles $\hol$ when $\Ra=0$.  $\hol$ should equal
 $\holp$ for vertical cylinders and $\holb$ for level cylinders.
 It was established earlier that when $\Ra=0$, cylinder $\hol$ is
 insensitive to orientation; let $\holz$ be the $\Ra=0$ surface
 conductance.  \tabref{tab:asymptotes} shows the desired coefficients
 for asymptotic values of $\holp$ and $\holb$:

\centerline{\bf\tabdef{tab:asymptotes}\quad{Cylinder asymptotic behaviors}}
\vbox{\settabs 5\columns
\toprule
\+\hfill $\vartheta~\qquad\qquad$ &\hfil$\holp$ &\hfil$\holb$ &\hfil$\holp$\bf~coefficient&\hfil$\holb$\bf~coefficient&\cr
\midrule
\+\hfill $90^\circ$\qquad\qquad &\hfil $\infty$&\hfil $\infty$ &\hfil 1 &\hfil 1 &\cr
\+\hfill $0^\circ$\qquad\qquad &\hfil $\infty$&\hfil $\infty$ &\hfil 1 &\hfil 1 &\cr
\+\hfill $90^\circ$\qquad\qquad &\hfil $\infty$&\hfil $\holz$ &\hfil 1 &\hfil 0 &\cr
\+\hfill $0^\circ$\qquad\qquad &\hfil $\holz$&\hfil $\infty$ &\hfil 0 &\hfil 1 &\cr
\+\hfill &\hfil &\hfil &\hfil $1-\holz/\holp$ &\hfil $1-\holz/\holb$ &\cr
\midrule
\+\hfill $90^\circ$\qquad\qquad &\hfil $\holz$&\hfil $\holz$ &\hfil 1 &\hfil 0 &\cr
\+\hfill $0^\circ$\qquad\qquad &\hfil $\holz$&\hfil $\holz$ &\hfil 0 &\hfil 1 &\cr
\+\hfill &\hfil &\hfil &\hfil $\sin^2\vartheta\,\holz^2\Big/\left[\holp\,\holb\right]$ &\hfil $\cos^2\vartheta\,\holz^2\Big/\left[\holp\,\holb\right]$ &\cr
\bottomrule
}
\medskip

 Combining the coefficients from \tabref{tab:asymptotes},
 Formula~\eqref{eq:inclined-cylinder-natural} satisfies these
 constraints.  Plotted in \figref{fig:conduction}, its error relative
 to $\holb$ Formula~\eqref{eq:level-cylinder} ($\vartheta=0^\circ$) is
 negligible; its error relative to $\holp$
 Formula~\eqref{eq:vertical-cylinder} ($\vartheta=90^\circ$) is less
 than 0.4\% when $H/d\ge1/9$.
$$\eqalign{\hol&=\left\|{\holp\left[1-{\holz\over\holp}+{\holz^2\,\sin^2\vartheta\over\holb\,\holp}\right]},~
                        {\holb\left[1-{\holz\over\holb}+{\holz^2\,\cos^2\vartheta\over\holb\,\holp}\right]}\right\|_{1+H/d}\cr
               &=\left\|{\holp-{\holz}+{\left[\holz\,\sin\vartheta\right]^2\over\holb}},~
                        {\holb-{\holz}+{\left[\holz\,\cos\vartheta\right]^2\over\holp}}\right\|_{1+H/d}}
  \eqdef{eq:inclined-cylinder-natural}$$

\medskip
\vbox{\settabs 1\columns
\+\hfil\figscale{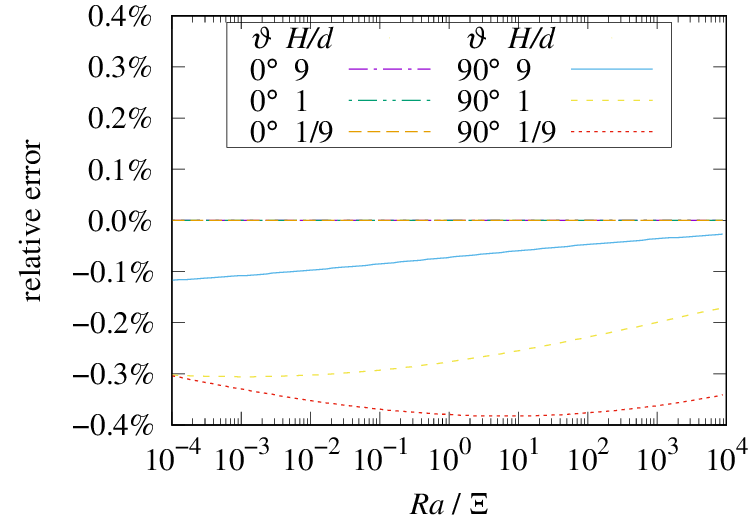}{234pt}&\cr
\+\hfil{\bf\figdef{fig:conduction}\quad{Cylinder at small $\Ra/\Xi$}}&\cr
}

\subsection{Heo and Chung}
 Heo and Chung~\cite{HEO2012366} measured copper electroplating onto a
 copper cylinder in a $\rm{CuSO_4/H_2SO_4}$ solution.  They measured
 the mass transfer coefficient~$\hol_m$, but reported their results as
 $\Nuol_d$ and $\Nuol_H$, which scale with different characteristic
 lengths.  The value of the mass transfer analog of $k$ was not reported, but for the purposes of
 comparing theory and measurements, $k$ is arbitrary if all
 conversions from $\Nuol$ to $\hol$ use the same~$k$.  A value of
 $k=1\rm{~mW/(m\cdot{K})}$ is used in \figref{fig:Heo-wide}.

\medskip
\vbox{\settabs 1\columns
\+\hfil\figscale{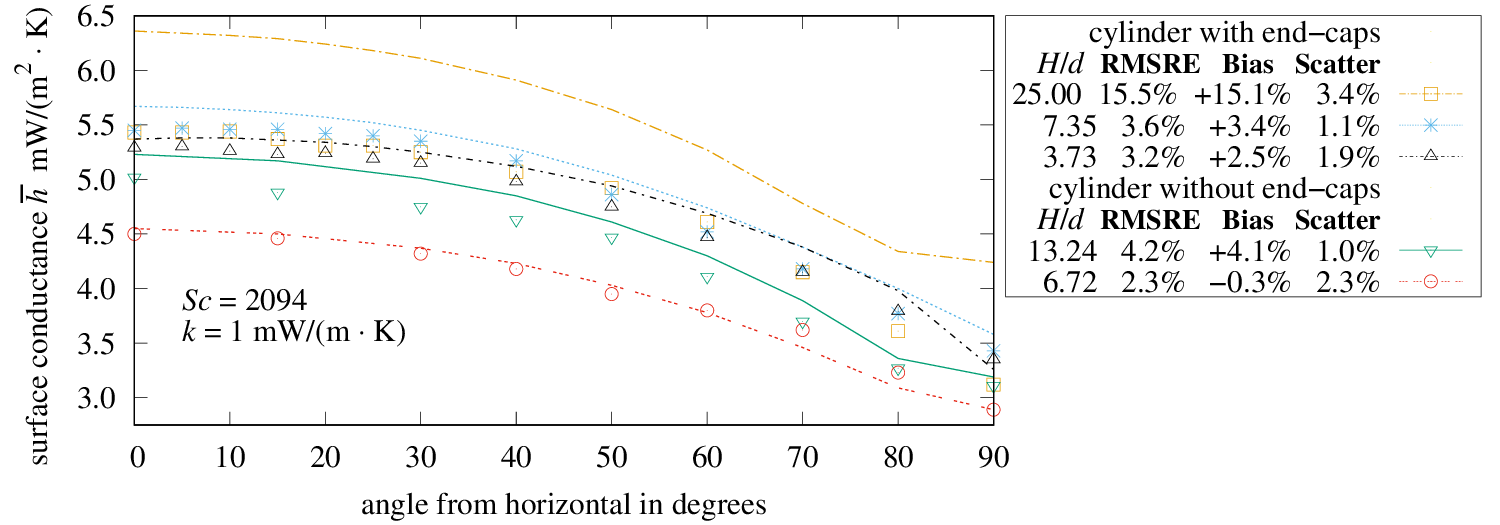}{468pt}&\cr
\+\hfil{\bf\figdef{fig:Heo-wide}\quad{Heo and Chung natural convection heat transfer from inclined cylinder}}&\cr
}
\medskip

 To avoid confusion, this section uses cylinder conductance
 $\Uol=\pi\,d\,H\,\hol$ with units W/K instead of~$\hol$.

 The cylinder and end-caps have different areas; their heat transfers
 must combine as conductances (W/K).  The top three rows
 in \figref{fig:Heo-wide} were modeled as:
$$\hol={\Uol+{\pi\,d^2}\,[\huol+\hdol]/4\over{\pi\,[d\,H+d^2/2]}}\eqdef{eq:U-with-end-caps}$$

 For the longer cylinders of the bottom two rows, $\hol$ should be
 nearly the same as the shorter cylinders; but the measured $\hol$
 values for $H/d=13.24$ and $H/d=6.72$ are significantly smaller
 in \figref{fig:Heo-wide}.  Modeling only the cylinder heat transfer,
 but including its end-cap areas yields RMSRE$\,<4.3\%$:
$$\hol={\Uol\over\pi\,[d\,H+d^2/2]}\eqdef{eq:h-Heo}$$

 Using their reported $\Pra=2094$, the $\Ra_d=5.07\times10^{10}$ in
 their table differs from the $\Ra_d=4.96\times10^{10}$ and
 $\Gr_d=2.37\times10^{7}$ in their figure.  $\Ra_d=5.07\times10^{10}$
 and $\Gr_d=2.42\times10^{7}$ are used in the $H/d=3.73$ and
 $H/d=6.72$ curves in \figref{fig:Heo-wide}.

 The parameters regarding the top row are more troubling.  Their table
 lists $\Ra_d=1.69\times10^8$ for the $d=0.010$~m cylinder.  Using
 their reported $\Pra=2094$ should result in
 $\Gr_d=\Ra_d/\Pra\approx8.07\times10^4$.  But their figures specify
 $\Gr_d=6.27\times10^4$ and $\Ra_d=1.31\times10^8$ for the $d=0.010$~m
 cylinder.  Given these inconsistencies,
 the $d=0.010$~m cylinder is omitted from the present work's summary
 statistics.

\medskip
\vbox{\settabs 1\columns
\+\hfil\figscale{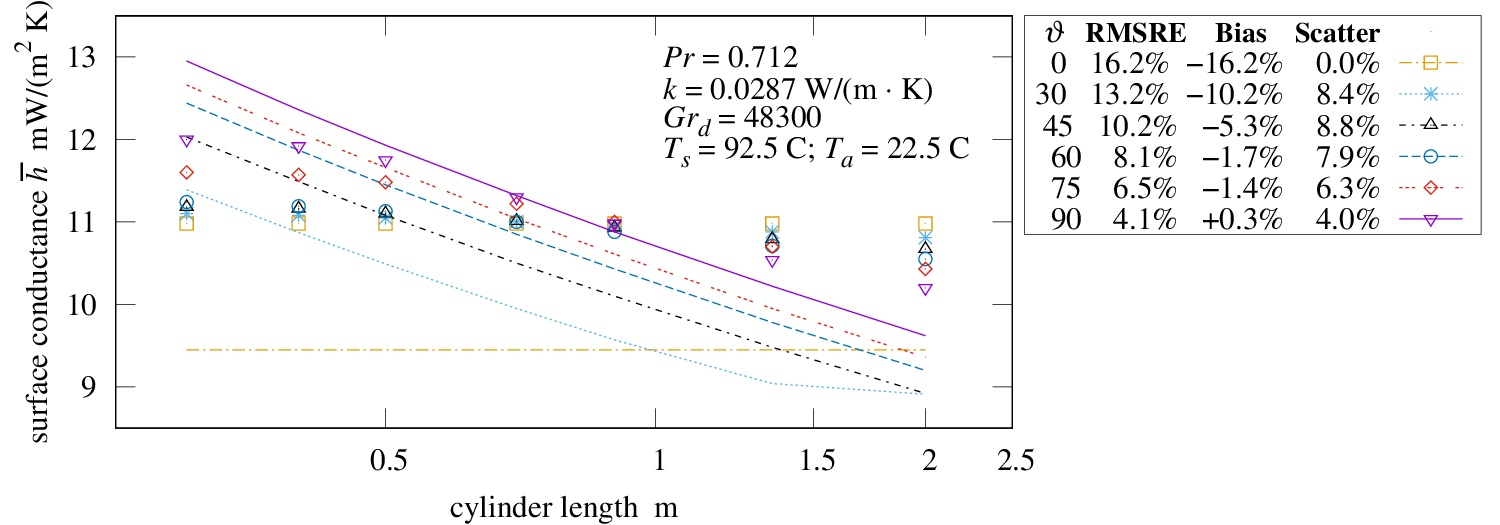}{468pt}&\cr
\+\hfil{\bf\figdef{fig:AlArabi-wide}\quad{Al-Arabi and Khamis natural convection heat transfer from inclined cylinder}}&\cr
}
\medskip

\subsection{Al-Arabi and Khamis}
 Al-Arabi and Khamis~\cite{ALARABI19823} measured local heat transfer
 to air from a ``nickel-electro-plated'' brass cylinder heated by
 steam.

 Text in their figures declares $\Gr_d=2.6\times10^4$; however, this
 is significantly smaller than the $\Gr_d=4.83\times10^4$ value this
 investigation computes from the average ambient conditions of Cairo,
 Egypt.  The present work uses $\Gr_d=4.83\times10^4$.

 For the average thermal surface conductance they report $h_L$ values
 instead of $\hol$, indicating that these are a local surface
 conductances, not average.  An earlier paper, Al-Arabi and
 Salman~\cite{ALARABI198045} reports local $h_L$ and $\hol$ values,
 and claims that $h_L$ and $\hol$ are ``practically the same''.  Yet
 this is clearly contradicted by the second figure of that paper.  The
 only angle for which they are the same is $\vartheta=90^\circ$.

 In vertical cylinder natural convection,
 growth of characteristic length $L$ corresponds to growth in the
 direction of fluid flow.  In such systems the average heat transfer
 can be inferred by averaging local heat transfers $h_x$ at
 lengths~$0<x\le{L}$.

 This fails for a horizontal cylinder because its
 characteristic-length is the cylinder's diameter, not its
 length.  \figref{fig:AlArabi-wide} averages the $h_x$ values to produce
 $\hol$, and confirms that this averaging works only for the vertical
 cylinder~$\vartheta=90^\circ$.  Only the vertical cylinder is
 included in the present work's summary statistics.

\medskip
\vbox{\settabs 1\columns
\+\hfil\figscale{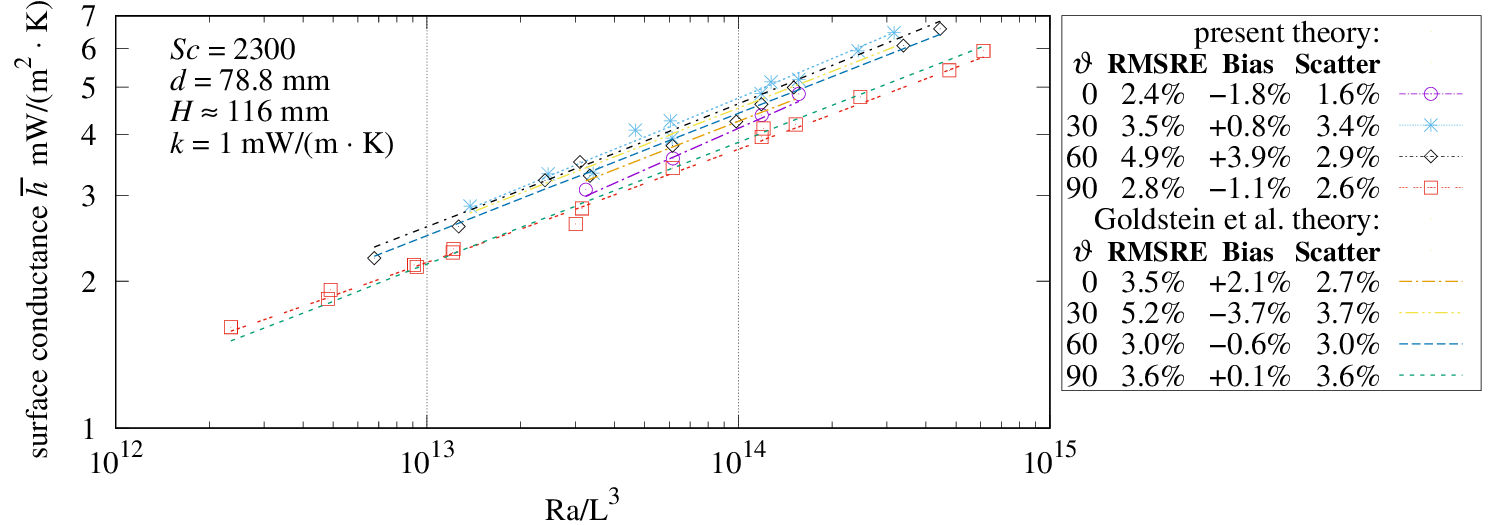}{468pt}&\cr
\+\hfil{\bf\figdef{fig:Goldstein-wide}\quad{Goldstein et al. natural convection heat transfer from inclined cylinder}}&\cr
}

\subsection{Goldstein et al.}
 Goldstein \etal\cite{GOLDSTEIN2007741} measured copper electroplating
 onto three cylinders at four angles in a $\rm{CuSO_4/H_2SO_4}$
 solution.  The cylinders were 78.8~mm in diameter and had lengths
 49.9~mm, 116.4~mm, and 184.4~mm.  They presented measurements without
 identifying the cylinder used in each trial.  The present analysis
 treats all as having length 116.4~mm, which succeeds because the
 present work $0.63\le{H/d}\le2.34$ traces
 in \figref{fig:natural-vertical} are converging above
 $\Ra/\Xi_\bullet>10^8$.

 The value of the mass transfer analog of $k$ was not reported, but for the purposes of
 comparing theory and measurements, $k$ is arbitrary if all conversions
 from $\Nuol$ to $\hol$ use the same~$k$.  A value of
 $k=1\rm{~mW/(m\cdot{K})}$ is used
 in \figref{fig:Goldstein-wide}.

\section{Discussion}

 Using the thermodynamics-based analysis pioneered by
 Jaffer~\cite{thermo3010010}, this investigation derived novel
 Formulas~(\eqrefn{eq:level-cylinder}, \eqrefn{eq:vertical-cylinder}, \eqrefn{eq:inclined-cylinder-natural})
 predicting the natural convective heat transfer from level, vertical,
 and inclined cylinders, respectively, given length $H$, diameter $d$,
 inclination angle $\vartheta$, $\Ra_d$, and the fluid's $\Pra$
 and~$k$, where $H/d\ge1/9$.

 These formulas enable the direct calculation of cylinder convective
 heat-transfer estimates at any inclination, avoiding the need for
 measurements of experimental prototypes or finite-element
 computations.

 End-cap heat-transfer Formulas~\eqref{eq:end-caps} were also proposed
 and tested on two of the Heo and Chung~\cite{HEO2012366} data-sets,
 yielding RMSRE of 3.2\% and 3.6\%.

\medskip
\centerline{\bf\tabdef{tab:convection-conformance}\quad{Measurements versus present theory}}
\vbox{\settabs 9\columns
\toprule
\+\bf Source && \hfil$\Pra$ or $\Sc$ &\hfil $H/d$&\hfil $\vartheta$ &\hfill\bf RMSRE&\hfill\bf Bias &\hfill\bf Scatter&\hfill\bf\#~&\cr
\midrule
\+Heo \& Chung \cite{HEO2012366} &&\hfil $Sc=2094$ &\hfil $7.4$ &\hfil $0^\circ$--$90^\circ$&\hfill $ 3.6\%$ &\hfill $ +3.4\%$ &\hfill $ 1.1\%$ &\hfill 13~&\cr
\+Heo \& Chung \cite{HEO2012366} &&\hfil $Sc=2094$ &\hfil $3.7$ &\hfil $0^\circ$--$90^\circ$&\hfill $ 3.2\%$ &\hfill $ +2.5\%$ &\hfill $ 1.9\%$ &\hfill 13~&\cr
\+Heo \& Chung \cite{HEO2012366} &&\hfil $Sc=2094$ &\hfil $13$ &\hfil $0^\circ$--$90^\circ$&\hfill $ 4.2\%$ &\hfill $ +4.1\%$ &\hfill $ 1.0\%$ &\hfill  9~&\cr
\+Heo \& Chung \cite{HEO2012366} &&\hfil $Sc=2094$ &\hfil $6.7$ &\hfil $0^\circ$--$90^\circ$&\hfill $ 2.3\%$ &\hfill $ -0.3\%$ &\hfill $ 2.3\%$ &\hfill  9~&\cr

\midrule
\+Nakai \& Okazaki \cite{SEIICHI1975387} &&\hfil $Pr=0.72$ & 10000--12500 &\hfil $0^\circ$&\hfill $ 1.5\%$ &\hfill $ -1.1\%$ &\hfill $ 1.0\%$ &\hfill 23~&\cr

\midrule
\+Goldstein et al.~\cite{GOLDSTEIN2007741} &&\hfil $Sc=2300$ &\hfil 0.63--2.34 &\hfil $ 0^\circ$&\hfill $ 2.4\%$ &\hfill $ -1.8\%$ &\hfill $ 1.6\%$ &\hfill  4~&\cr
\+Goldstein et al.~\cite{GOLDSTEIN2007741} &&\hfil $Sc=2300$ &\hfil 0.63--2.34 &\hfil $30^\circ$&\hfill $ 3.5\%$ &\hfill $ +0.8\%$ &\hfill $ 3.4\%$ &\hfill 11~&\cr
\+Goldstein et al.~\cite{GOLDSTEIN2007741} &&\hfil $Sc=2300$ &\hfil 0.63--2.34 &\hfil $60^\circ$&\hfill $ 4.9\%$ &\hfill $ +3.9\%$ &\hfill $ 2.9\%$ &\hfill 11~&\cr
\+Goldstein et al.~\cite{GOLDSTEIN2007741} &&\hfil $Sc=2300$ &\hfil 0.63--2.34 &\hfil $90^\circ$&\hfill $ 2.8\%$ &\hfill $ -1.1\%$ &\hfill $ 2.6\%$ &\hfill 16~&\cr

\midrule
\+AlArabi \& Khamis \cite{ALARABI19823} &&\hfil $Pr=0.708$ &\hfil 15.5--104 &\hfil $90^\circ$&\hfill $ 4.1\%$ &\hfill $ +0.3\%$ &\hfill $ 4.0\%$ &\hfill  7~&\cr

\bottomrule
}
\medskip

 \tabref{tab:convection-conformance} summarizes the present theory's
 conformance with 116 inclined cylinder measurements having
 $1.48<H/d<12500$ at angles $0^\circ\le\vartheta\le90^\circ$ in ten
 data-sets from four peer-reviewed studies, yielding (data-set) RMSRE
 values between 1.5\% and~5\%.

\medskip
\centerline{\bf\tabdef{tab:CC-3-conformance}\quad{Churchill and Chu versus theories}}
\vbox{\settabs 9\columns
\toprule
\+\bf Source &&\bf Theory&$\Ra_d/\Xi_\bullet\ge$ &~~$\Ra_d/\Xi_\bullet\le$&\hfill\bf RMSRE&\hfill\bf Bias &\hfill\bf Scatter&\hfill\bf\#~&\cr
\midrule
\+           Churchill \& Chu~\cite{CHURCHILL19751049} && turbulent& $7.5\times10^{-12}$ &~~$3.3\times10^{9}$&\hfill $21.1\%$ &\hfill $ +9.8\%$ &\hfill $18.7\%$ &\hfill 57~&\cr
\+           Churchill \& Chu~\cite{CHURCHILL19751049} && laminar& $7.5\times10^{-12}$ &~~$3.3\times10^{9}$&\hfill $19.0\%$ &\hfill $ -7.9\%$ &\hfill $17.2\%$ &\hfill 57~&\cr
\+           Churchill \& Chu~\cite{CHURCHILL19751049} && present& $7.5\times10^{-12}$ &~~$3.3\times10^{9}$&\hfill $11.4\%$ &\hfill $ -1.0\%$ &\hfill $11.3\%$ &\hfill 57~&\cr

\bottomrule
}
\medskip

 On 57 level cylinder measurements from Churchill and
 Chu~\cite{CHURCHILL19751049} in \tabref{tab:CC-3-conformance}
 spanning more than 20 orders-of-magnitude of $\Ra$, present
 Formula~\eqref{eq:level-cylinder} has 11\% RMSRE, a significant
 improvement from the 21\% and 19\% RMSRE of prior work
 Formulas~\eqref{eq:Nu-level-cylinder-CC}
 and \eqref{eq:Nu-level-cylinder-CC-lam}.

\subsection{Laminar and Turbulent Flows}
 Heo and Chung~\cite{HEO2012366} claimed that natural convection from
 four of their five cylinders was turbulent.
 Goldstein \etal\cite{GOLDSTEIN2007741} claimed that most of their
 data was for laminar natural convection, but that their single
 correlation was ``reasonably accurate'' on all their data.
 If they are correct about the flow modes induced by their cylinders,
 then the present Formula~\eqref{eq:inclined-cylinder-natural}
 $1.9\%\le{\rm RMSRE}\le4.8\%$ performance
 in \tabref{tab:convection-conformance} includes both laminar and
 turbulent natural convection.

 $\Ra_d$ in level cylinder Formula~\eqref{eq:level-cylinder} has
 neither the 1/4 exponent commonly attributed to laminar flow nor the
 1/3 exponent attributed to turbulent flow, but an intermediate
 exponent of $1/[2+E^\bullet]\approx0.310$.

\subsection{Local Convective Heat Transfer}
 The technique of calculating average heat transfer $\hol$ by
 averaging local heat transfers $h_x$ at lengths~$0<x\le{L}$ works
 only when $h_x$ is constant across the width of a plate or the
 circumference of the cylinder.  For natural convection, this is only
 when the plate or cylinder is vertical.

 Otherwise, each $h_{x,y}$ must be averaged, requiring a
 two-dimensional local convection model.  Lacking such a model, only
 the vertical cylinder data-set of
 Al-Arabi and Khamis~\cite{ALARABI19823} is valid.

 Thermodynamic constraints can be powerful tools, but apply only to
 complete systems.  While a thermodynamic constraint would be
 intrinsic to a perfect molecular simulation, such a simulation is
 impractical.  Furthermore, thermodynamic constraints might well not
 survive the truncation errors of digital computation.

\subsection{Short Cylinders}
 Around a short ($H\ll{d}$) level cylinder the fluid flow is not
 restricted to the vertical plane, resulting in larger heat transfers
 than predicted by Formula~\eqref{eq:inclined-cylinder-natural}.
 Formula~\eqref{eq:inclined-cylinder-natural} should work for small
 $H/d$ ratios as long as $H$ is the width of a heated band embedded in
 a longer, insulated cylinder as shown in \figref{fig:banded}.

\medskip
\vbox{\settabs 1\columns
\+\hfil\figscale{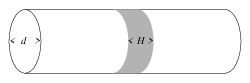}{234pt}&\cr
\+\hfil{\bf\figdef{fig:banded}\quad{Cylinder with isothermal band}}&\cr
}

\subsection{Non-Circular Cylinders}
 For a convex cylinder with hydraulic diameter $d$, vertical
 Formula~\eqref{eq:vertical-cylinder} is expected to predict heat
 transfer correctly.

 For a level convex cylinder, the \tabref{tab:cylinder convection
 parameters} natural convection parameters $B$, $D$, and $E$ need to
 be reevaluated.  $B$ should be the cross-section's perimeter length
 squared divided by its area.  $D$ will be less than $B$.  Parameter
 $E$ (average bend divided by $\pi/2$~rad) can be calculated by the
 method of
 Formulas~\eqref{eq:average-bottom-angle}, \eqref{eq:average-top-angle},
 and~\eqref{eq:E-bullet}.

 If the cross-section lacks bilateral symmetry, then the convection
 should be calculated separately for each side of the
 cross-section, split along the line connecting its highest and lowest
 point of the cross-section.

\subsection{Rough Cylinders}
 Jaffer and Jaffer~\cite{tse9275} made natural convective heat
 transfer measurements of a 0.305~m square plate with 3~mm
 root-mean-square height-of-roughness at angles between $-90^\circ$
 and $+90^\circ$.  Those measurements matched
 Formula~\eqref{eq:inclined-natural} with 3\%~RMSRE, providing
 evidence that flat surface natural convection is insensitive to
 roughness which is much smaller than its characteristic length.

 A similar test conducted with a rough cylinder would ascertain
 whether the natural convective flows from cylinders are also
 insensitive.

\beginsection{Acknowledgments}

 Thanks to John H. Lienhard V and anonymous reviewers for their useful
 suggestions.

\section{Nomenclature}

\nomenclature[A]{$B, C, D, E$}{dimensionless natural convection parameters}
\nomenclature[A]{$E^\bullet$}{exponent parameter for level cylinders}
\nomenclature[A]{$H$}{cylinder length (m)}
\nomenclature[A]{$d$}{cylinder diameter (m)}
\nomenclature[A]{$h_x$}{local convective surface conductance (${\rm W/(m^2\cdot K)}$)}
\nomenclature[A]{$\hol$}{average convective surface conductance (${\rm W/(m^2\cdot K)}$)}
\nomenclature[A]{$\hsol$}{upward convective surface conductance (${\rm W/(m^2\cdot K)}$)}
\nomenclature[A]{$\hqol$}{vertical plate convective surface conductance (${\rm W/(m^2\cdot K)}$)}
\nomenclature[A]{$\hRol$}{downward convective surface conductance (${\rm W/(m^2\cdot K)}$)}
\nomenclature[A]{$\holb$}{level cylinder convective surface conductance (${\rm W/(m^2\cdot K)}$)}
\nomenclature[A]{$\holp$}{vertical cylinder convective surface conductance (${\rm W/(m^2\cdot K)}$)}
\nomenclature[A]{$\holz$}{$\Ra=0$ cylinder conductive surface conductance (${\rm W/(m^2\cdot K)}$)}
\nomenclature[A]{$\huol$}{upper end-cap convective surface conductance (${\rm W/(m^2\cdot K)}$)}
\nomenclature[A]{$\hdol$}{lower end-cap convective surface conductance (${\rm W/(m^2\cdot K)}$)}

\nomenclature[A]{$k$}{fluid thermal conductivity (${\rm W/(m\cdot K)}$)}
\nomenclature[A]{$L$}{characteristic length (m)}
\nomenclature[A]{$\Ls$}{characteristic length of upward-facing surface (m)}
\nomenclature[A]{$L'$}{characteristic length of vertical surface (m)}
\nomenclature[A]{$L_R$}{characteristic length of downward-facing surface (m)}

\nomenclature[A]{$\Nuzq$}{Nusselt number of vertical plate conduction}
\nomenclature[A]{$\Nuzs$}{Nusselt number of upward-facing plate conduction}
\nomenclature[A]{$\Nuzb$}{Nusselt number of level cylinder conduction}
\nomenclature[A]{$\Nuol$}{average Nusselt number}
\nomenclature[A]{$\Nuols$}{Nusselt number of upward-facing plate}
\nomenclature[A]{$\Nuolq$}{Nusselt number of vertical plate plate}
\nomenclature[A]{$\NuolR$}{Nusselt number of downward-facing plate}
\nomenclature[A]{$\Nuolb$}{Nusselt number of level cylinder}
\nomenclature[A]{$\Nuolp$}{Nusselt number of vertical cylinder}

\nomenclature[A]{$p$}{exponent in $\ell^{p}$-norm}
\nomenclature[A]{$\Pra$}{Prandtl number of the fluid}
\nomenclature[A]{$\Ra$}{Rayleigh number}
\nomenclature[A]{$\Ra_d$}{Rayleigh number with cylinder diameter as characteristic length}
\nomenclature[A]{$\Ra_H$}{Rayleigh number with cylinder length as characteristic length}
\nomenclature[A]{$\Ra^*$}{upward Rayleigh number with characteristic length $\Ls$}
\nomenclature[A]{$\Ra'$}{vertical plate Rayleigh number with characteristic length $L'$}
\nomenclature[A]{$\Ra_R$}{downward Rayleigh number with characteristic length $L_R$}

\nomenclature[A]{$\Sc$}{Schmidt number of the fluid}
\nomenclature[A]{$\Shol$}{average Sherwood number}
\nomenclature[A]{$U_0$}{thermal conductance of isothermal sphere in uniform medium (W/K)}
\nomenclature[A]{$\Uol$}{convective thermal conductance of cylinder (W/K)}

\subsection{Greek Symbols}
\unskip

\nomenclature[G]{$\theta$}{angle of the plate from vertical}
\nomenclature[G]{$\vartheta$}{angle of the cylinder axis from horizontal}
\nomenclature[G]{$\Xi$}{self-obstruction factor for plates and vertical cylinders}
\nomenclature[G]{$\Xi_\bullet$}{self-obstruction factor for level cylinders}

\section{References}

\bibliographystyle{unsrtDOI}
\bibliography{citations}

\bye